\documentclass[preprintnumbers,secnumarabic,amsmath,amssymb,nofootinbib,floatfix]{revtex4}
\usepackage{varioref,exscale,latexsym,amsmath,amssymb}
\usepackage{graphicx}

\usepackage{graphicx}
\usepackage{slashed}
\usepackage{dcolumn}
\usepackage{bm}
\usepackage{hyperref}
\usepackage{color}
\usepackage{xcolor}
\newcommand{\beq}{\begin{equation}}
\newcommand{\eeq}{\end{equation}}
\newcommand{\bea}{\begin{eqnarray}}
\newcommand{\eea}{\end{eqnarray}}
\newcommand{\nn}{\nonumber}

\def\lsi{\raise0.3ex\hbox{$<$\kern-0.75em\raise-1.1ex\hbox{$\sim$}}}
\def\gsi{\raise0.3ex\hbox{$>$\kern-0.75em\raise-1.1ex\hbox{$\sim$}}}

\def\beq{\begin{equation}}

\def\eeq{\end{equation}}

\def\beqa{\begin{eqnarray}}

\def\eeqa{\end{eqnarray}}

\begin{document}
\preprint{ACFI-T18-06}

\title{{\bf Gauge Assisted Quadratic Gravity: \\
A Framework for UV Complete Quantum Gravity }}

\medskip\

\medskip\

\author{John F. Donoghue}
\email{donoghue@physics.umass.edu}
\affiliation{~\\
Department of Physics,
University of Massachusetts\\
Amherst, MA  01003, USA\\
 }

\author{Gabriel Menezes}
\email{gsantosmenez@umass.edu}
\affiliation{~\\
Department of Physics,
University of Massachusetts\\
Amherst, MA  01003, USA\\
}

\affiliation{~ Departamento de F\'{i}sica, Universidade Federal Rural do Rio de Janeiro, 23897-000, Serop\'{e}dica, RJ, Brazil \\
 }

\begin{abstract}
We discuss a variation of quadratic gravity in which the gravitational interaction remains weakly coupled at all energies, but is assisted by a Yang-Mills gauge theory which becomes strong at the Planck scale. The Yang-Mills interaction is used to induce the usual Einstein-Hilbert term, which was taken to be small or absent in the original action. We study the spin-two propagator in detail, with a focus on the high mass resonance which is shifted off the real axis by the coupling to real decay channels. We calculate scattering in the $J=2$ partial wave and show explicitly that unitarity is satisfied. The theory will in general have a large cosmological constant and we study possible solutions to this, including a unimodular version of the theory. Overall, the theory satisfies our present tests for being a ultraviolet completion of quantum gravity.
\end{abstract}
\maketitle

\section{Introduction}

There are many exotic approaches to quantum gravity, but comparatively little present work exploring the option of describing gravity by a renormalizable quantum field theory (QFT). Nature has shown that the other fundamental interactions, i.e. those of the Standard Model, are described by renormalizable QFTs at present energies, so this should be the most conservative approach. There is a modest body of recent work
\cite{Einhorn, Strumia, Donoghue, DM, Holdom, Mannheim, Hooft, Shapiro, Tomboulis, Smilga, Narain, Ans,Sobreiro:12} attempting to
revive this possibility\footnote{For the much larger body of older research please see the references within \cite{Einhorn, Strumia, Donoghue, DM, Holdom, Mannheim, Hooft, Shapiro, Tomboulis, Smilga, Narain, Ans,Sobreiro:12}.}. As a subfield, this literature is somewhat diffuse, with different groups pursuing different, although related, variants. However, the key question is to determine if {\em any} renormalizable QFT can serve as a UV completion for gravity. In this paper, we explore such a variation which we feel is particularly well suited for being a controlled approach using present techniques, with encouraging results.

The distinctive feature of a renormalizable QFT treatment of gravity is simple to diagnose. Loops involving matter fields coupled to the metric yield divergences proportional to the second power of the curvatures. Therefore, the fundamental action must have terms in it which are quadratic in the curvatures in order to renormalize the theory. This also explains why such QFT treatments are often overlooked. Curvatures involve second derivatives of the metric, so that quadratic gravity involves metric propagators which are quartic in the momentum. Quartic propagators are generally considered problematic, for reasons which will be reviewed below. However, quadratic gravity does have the positive feature that it is renormalizable \cite{Stelle:1976gc}, and can be asymptotically free \cite{Julve:1978xn, Fradkin:1981hx, Fradkin:1981iu}. Moreover, to recover General Relativity in the low energy limit one must arrange to have the usual quadratic propagators at low energy. The challenge for such QFT treatments then is to deal with fundamental quartic propagators at high energy while recovering usual gravity at low energy.

The variation which we explore will involve a Yang-Mill gauge theory plus weakly coupled quadratic gravity. For the purposes of this introduction one can consider this to be defined by the action (in units of  $\hbar = c = k_{B} = 1$, which will be consistently employed throughout the paper)
\begin{equation}
S= \int d^4x \sqrt{-g} \left[-\frac{1}{4g^2}g^{\mu\alpha}g^{\nu\beta} F_{\mu\nu}^a F_{\alpha\beta}^a  - \frac{1}{2\xi^2} C_{\mu\nu\alpha\beta}C^{\mu\nu\alpha\beta} \right]
\label{action1}
\end{equation}
although we also discuss other variants below. The Weyl tensor is given by
\begin{eqnarray}
C_{\mu\nu\alpha\beta} &=& {R}_{\mu\nu\alpha\beta} -\frac12 \left( {R}_{\mu\alpha} g_{\nu\beta} - {R}_{\nu\alpha} g_{\mu\beta} - {R}_{\mu\beta} g_{\mu\alpha} + {R}_{\nu\beta} g_{\mu\alpha} \right) \ \ \nonumber \\
&+& \frac{{R(g)}}{6}\left(g_{\mu\alpha} g_{\nu\beta} - g_{\nu\alpha} g_{\mu\beta} \right)  \ \ .
\end{eqnarray}
The quantity $F^{a}_{\mu\nu}$ is the usual field strength tensor for the Yang-Mills theory. The index $a$ is summed over the generators of the gauge group $G$. One has that
\beq
F^{a}_{\mu\nu} = \partial_{\mu} A^{a}_{\nu} - \partial_{\nu} A^{a}_{\mu} + g f^{abc} A^{b}_{\mu} A^{c}_{\nu}
\eeq
where $f^{abc}$ are the structure constants of $G$ and $g$ is the Yang-Mills coupling constant.

Both couplings $g$ and $\xi$ are asymptotically free, so this is a renormalizable asymptotically free theory at high energy. We consider the limit where the gauge coupling $g$ is larger than the gravitational coupling $\xi$. This means that the gauge theory becomes strongly interacting at a higher energy. The energy scale of the gauge theory will be taken to be the Planck scale. Indeed, this is the role of the helper gauge theory in this construction - it defines the Planck scale, and also induces the Einstein-Hilbert action at lower energies \cite{Adler:1980pg, Zee:1980sj, Zee:1981mk, Adler:1982ri, Brown:1982am}. We have elsewhere \cite{DM} calculated the induced gravitational constant due to QCD,
\begin{equation}
\frac{1}{16\pi G_{\textrm{ind}}} = 0.0095 \pm 0.0030~ {\rm GeV}^2
\end{equation}
and so a QCD-like theory would need an energy scale $10^{19}$ times greater to generate the observed Planck scale.

Left on its own, the Weyl-squared term would also become strong at some energy. However, because we take it to be still weakly coupled at the Planck scale, it would become strong only at a very low energy scale. (For example, for a pure Weyl theory if $\xi =0.1$ at the Planck scale, it would become strong at $\Lambda_{\xi} = 10^{-1006}$~eV.) In practice, this running of $\xi$ is interrupted by the induced gravitational effects due to the gauge field, and hence is never relevant at low energy\footnote{The same thing happens to the SU(2) coupling of the Standard Model. It would become strong at $10^{-14}$ eV but its running is interrupted by the symmetry breaking at the TeV scale.}. Indeed we will see that {\em the gravitational interaction can remain weakly coupled at all energies.} This helps give us control over the predictions of the theory. Our goal is to explore the structure of this theory in order to see if there are any calculable obstacles to treating it as a UV completion of quantum gravity. The theory so far passes such tests successfully.

In order to describe the nature of this theory, let us show a subset of our results. For the purpose of this introduction, let us again simplify the result by ignoring the possibility of an induced cosmological constant, although this will be discussed below. In this case, the spin-two part of the propagator will be seen to have the following structure
\begin{eqnarray}
iD_{\mu\nu\alpha\beta} &=& i{\cal P}^{(2)}_{\mu\nu\alpha\beta} D_2(q)
\nonumber \\
D^{-1}_2(q) &=& \frac{q^2+i\epsilon}{\tilde{\kappa}^2(q)}- \frac{q^4}{2\xi^2(\mu)}
- \frac{q^4 N_{\textrm{eff}}}{640\pi^2} \ln \left(\frac{-q^2-i\epsilon}{\mu^2}\right)
- \frac{q^4 N_{q}}{1280\pi^2} \ln \left[\frac{(q^2)^2}{\mu^4}\right]~
\end{eqnarray}
Here, ${\cal P}^{(2)}_{\mu\nu\alpha\beta}$ is the spin-two projector to be described below. The function $1/\tilde{\kappa}^2(q)$ is induced by the gauge theory and can be described by techniques related to QCD sum rules. It will have the limits
\begin{eqnarray}
\frac{1}{\tilde{\kappa}^2(q)} &\to& \frac{1}{\kappa^2}~~~~~q \ll M_P \nonumber \\
&\to& 0~~~~~~q \gg M_P
\end{eqnarray}
where $\kappa^2 = 32\pi G$ is the coupling in the Einstein-Hilbert action. The coefficient of the first logarithm, $N_{\textrm{eff}}$, is a number that depends on the number of light degrees of freedom with the usual couplings to gravity and $N_q$ is a number due to gravitons coupled through the quadratic-curvature action. These numbers will be defined in detail below but for now we can note that at very high energies $N_{\textrm{eff}} = N_\infty = D + N_{SM}$, where $D$ is the number of generators in the gauge theory and $N_{SM}$ is due to the particles of the Standard Model and beyond\footnote{Including just the Standard Model particles yields $N_{SM} = 283/12$.} while $N_q=199/3$.

This propagator is described by three regions. At low energy $q^2 \ll \xi^2 M_P^2$, where $M_{p}$ is the Planck mass\footnote{To be more precise, we are referring to a variation on a reduced Planck mass $M_p^2 =2/\kappa^2$}, the quadratic propagator dominates, and one gets the usual coupling of General Relativity. At high energy $q^2> M_P^2$, one has a purely quartic propagator and the running gravitational coupling is
\begin{equation}
\xi^2 (q) = \frac{\xi^2(\mu) }{1+ \frac{ \xi^2(\mu) (N_{\infty}+N_q)}{320\pi^2} \ln (q^2/\mu^2)} = \frac{320\pi^2}{(N_{\infty}+N_q) \ln(q^2/\Lambda_{\xi}^2)}
\end{equation}
where as usual we have defined the scale factor $\Lambda_{\xi}$ via
$1/\xi^2(\mu) = (N_{\infty}+N_q) (\ln \mu^2/\Lambda_{\xi}^2)/320 \pi^2$. In the intermediate regime
$\xi^2 M_P^2 \leq q^2 <  M_P^2$, the propagator goes through a resonance, and also transitions from quadratic to quartic behavior. A plot of the absolute value of propagator throughout these regions is shown in Fig. 1 for time-like values of $q^2$, normalized to the usual propagator at low energy. One can readily see the usual behavior at low energy, as well as the improved momentum behavior at high energy. But clearly the striking feature is the resonance, which occurs at $q^2 = m_r^2= 2\xi^2/{\kappa}^2$. At weak coupling the width is roughly $\Delta q^2 \sim N_{\textrm{eff}} \xi^2  m_r^2/320\pi $. It is fair to think of this as an unstable spin-two resonance, and we will explore this interpretation more fully.

\begin{figure}[htb]
\begin{center}
\includegraphics[height=60mm,width=65mm]{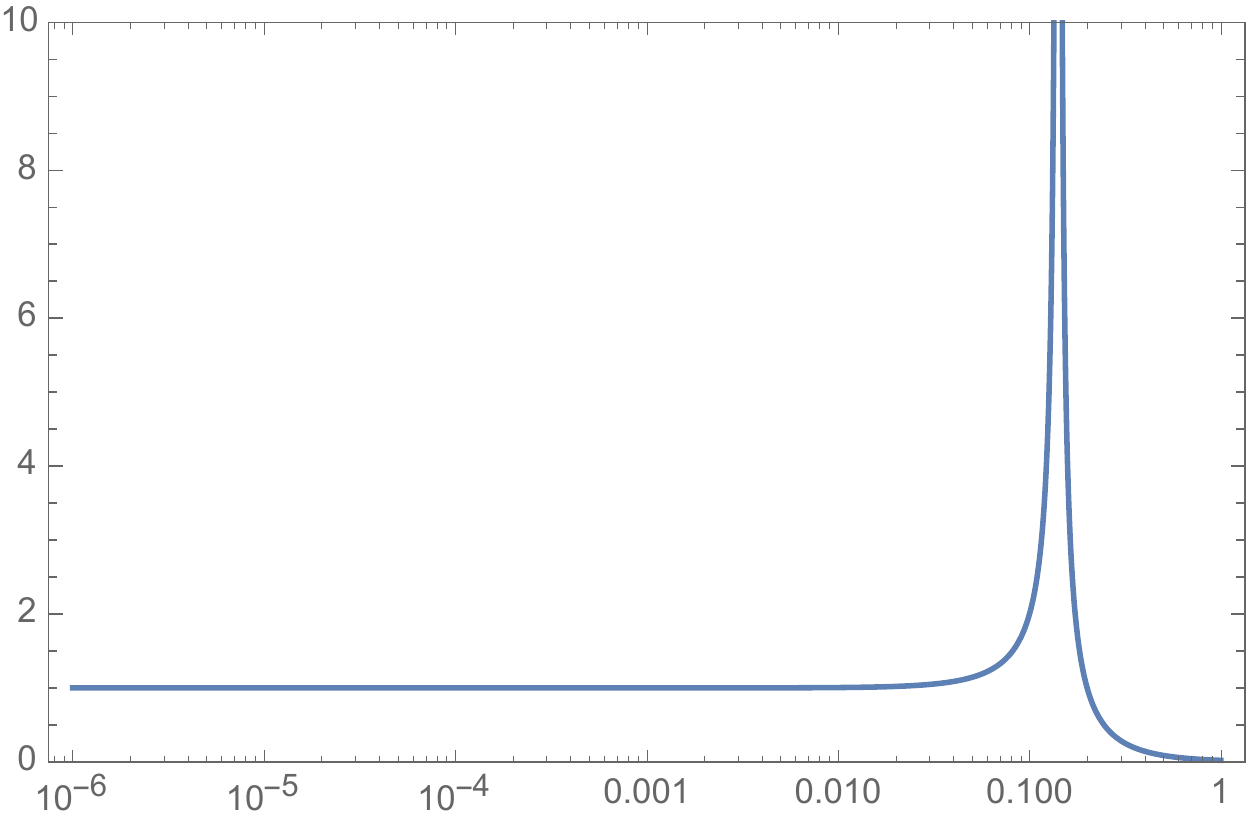}
\caption{The absolute value of the spin-two propagator for $\xi=0.1$, normalized to the standard propagator of General Relativity. The x-axis is the momentum $|q|$ in the time-like region, in units of the Planck mass. The imaginary parts have been calculated with loops of Standard Model particles and gravitons. }
\label{grec}
\end{center}
\end{figure}

We have used this propagator in the calculation of a physical process which goes through the resonance region, which is the scattering in the spin-two s-channel partial wave. The cross-section is shown in Fig. 2.  The usual growth of the amplitude, which normally would become strong at the Planck scale, is tamed by the quartic part of the propagator beyond $s=m_r^2$, and remains weakly coupled. We will explicitly confirm that the scattering amplitude is unitary at all energies. This occurs despite the sign of the propagator near the pole being the opposite from normal expectation. Careful readers should pay attention to this point below.

\begin{figure}[htb]
\begin{center}
\includegraphics[height=60mm,width=65mm]{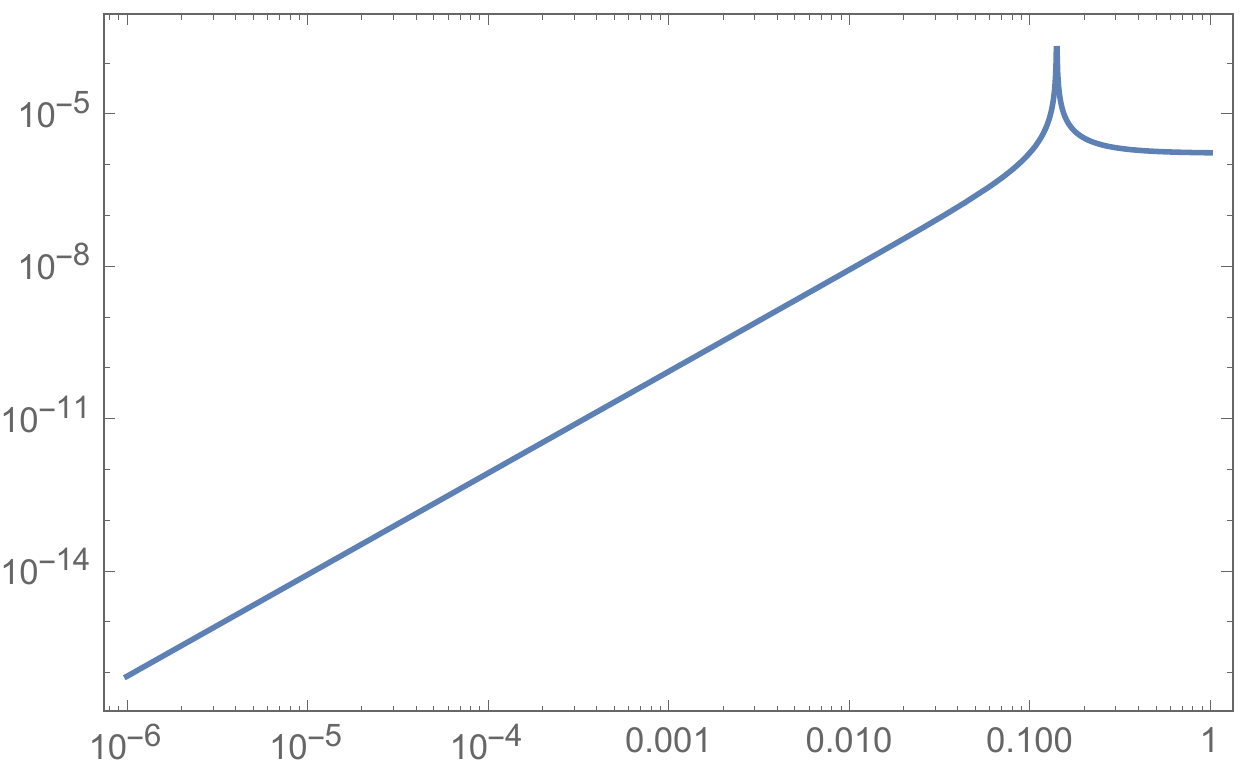}
\caption{The absolute value of the unitary $J=2$ partial wave amplitude, calculated with $\xi=0.1$. The x-axis is the center of mass energy in units of the Planck mass.}
\label{grec2}
\end{center}
\end{figure}

There are some approximations which have gone into the form of the propagator. One is that we have included the logarithmic behavior from loops but not any residual non-logarithmic constants. These in practice can be absorbed in a redefinition of $1/\xi^2$, but we have not explicitly calculated these terms. More importantly, we have treated each region using only the dominant action appropriate for that region. In particular this means that when we are displaying results from the highest energies, we use the quadratic curvature action only. This feature is related to the lack of an imaginary part in the $\ln q^4$ terms.

We also note that the propagator in the space-like region, and equally the Euclidean propagator, is featureless and well behaved. In particular the  Euclidean propagator involves
\begin{eqnarray}
D_{2E}(q_E) &=& \left[\frac{q_E^2}{\tilde{\kappa}^2(q_E)}+\frac{q_E^4}{2\xi^2(\mu)}+ \frac{q_E^4 (N_{\textrm{eff}} + N_{q})}{640\pi^2} \ln \left(\frac{q_E^2}{\mu^2}\right)\right]^{-1}
\nn\\
&=& \left[\frac{q_E^2}{\tilde{\kappa}^2(q_E)} +\frac{q_E^4 (N_{\textrm{eff}} + N_{q})}{640\pi^2}
 \ln \left(\frac{q^2_E}{\Lambda^2_{\xi}}\right)\right]^{-1}
\end{eqnarray}
where $q^2_E>0$. It is shown in Fig. 3, again normalized to the usual quadratic propagator. Note that without the helper gauge interaction inducing the usual gravitational coupling at low energy, the quartic terms would have blown up at $q_E^2 = \Lambda_{\xi}^2$ but this point is innocuous in the present framework.

\begin{figure}[htb]
\begin{center}
\includegraphics[height=60mm,width=65mm]{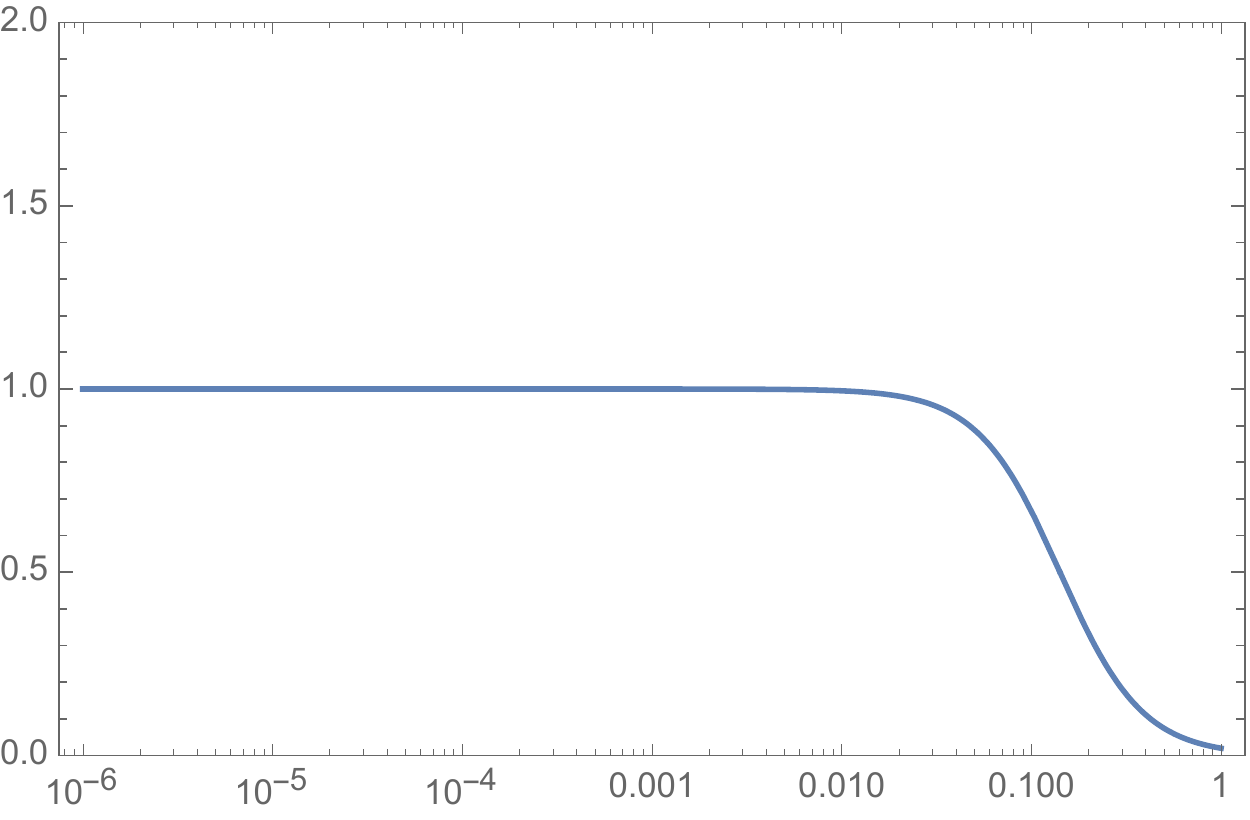}
\caption{The Euclidean spin-two propagator corresponding to the same conditions as shown in Fig. 1, again normalized to the standard propagator of General Relativity. }
\label{grec3}
\end{center}
\end{figure}

With this as introduction, we turn to the more detailed investigation of the theory. In Sec. 2 we describe some of the issues of quadratic gravity in general, including some aspects of describing the degrees of freedom. In Sec. 3 we turn to the spin-two propagator and explore some of the results mentioned above. Section 4 describes a test of unitarity in scattering in the spin-two channel. Sec 5 gives a discussion of a broader class of theories, including the cosmological constant and also the spin-zero sector. In Sec. 6, we make some comments on the unimodular version of this theory, which also may be useful to explore more fully. Finally Sec. 7 gives a summary and discussion. In the Appendix, we calculate the magnitude of the $R^2$ term in the action which is also induced by the Yang-Mills interaction.

\section{Quadratic Gravity}

In addition to the Einstein-Hilbert action
\begin{equation}
S_{\textrm{EH}} = \frac{2}{\kappa^2}\int d^4 x \sqrt{-g} R
\end{equation}
there in general can be three combinations at quadratic order in the curvature which are generally covariant
\begin{equation}\label{quadraticorder}
S_{\textrm{quad}} = \int d^4x \sqrt{-g}
\left[\frac{1}{6 f_0^2} R^2 - \frac{1}{2\xi^2} C_{\mu\nu\alpha\beta}C^{\mu\nu\alpha\beta} - \eta G\right]
\end{equation}
where
\begin{equation}
G = R_{\mu\nu\alpha\beta} R^{\mu\nu\alpha\beta} - 4 R_{\mu\nu} R^{\mu\nu} + R^2
\end{equation}
is the Gauss-Bonnet invariant. This latter term is a total derivative in four dimensions, and so it cannot influence the classical equations of motion nor graviton propagation. One can also introduce a surface term $\Box R$ in the above action. The counterterm associated with it in the calculation of the one-loop effective action is gauge-dependent. We will drop the surface term as well as the Gauss-Bonnet contribution in the rest of this paper. In any case, we remark that topological and surface terms should be included in order to provide renormalizability.
We also note that
\beq \label{weylidentity}
- \frac{1}{2\xi^2} C_{\mu\nu\alpha\beta}C^{\mu\nu\alpha\beta} =  -\frac{1}{\xi^2}\left[\left(R_{\mu\nu}R^{\mu\nu}
- \frac13 R^2\right) +\frac{1}{2} G\right]
\eeq
Our conventions are: the Minkowski metric is given as $\eta_{\mu\nu} = \textrm{diag}(1,-1,-1,-1)$ and the Riemann curvature tensor is given by $R^{\lambda}_{\ \mu\nu\kappa} = \partial_{\kappa}\Gamma^{\lambda}_{\mu\nu} + \Gamma^{\eta}_{\mu\nu}\Gamma^{\lambda}_{\kappa\eta} - (\nu \leftrightarrow \kappa)$.

In theories with fundamental curvature-squared terms, the graviton propagator will be quartic in the momentum. This is generally considered to be problematic. With a quartic propagator in free field theory one expects negative norm ghost states, using for example
\begin{equation}\label{quartic}
\frac{-i}{q^4} \sim \frac{-i}{q^2(q^2-\mu^2)} = \frac{1}{\mu^2}\left(\frac{i}{q^2}- \frac{i}{q^2-\mu^2}\right)
\end{equation}
and perhaps tachyons if $\mu^2$ is negative. When considering quadratic curvature gravity, the decomposition of the degrees of freedom varies depending on the gauge condition imposed and also on the choice of field parametrization, and there are generally gauge-variant unphysical states~\cite{bos}. However, it is generally agreed that the free field propagating modes are a massive scalar and its massless scalar ghost, and a massless spin-two graviton and its massive spin-two ghost. The scalar massive mode arises uniquely from the $R^2$ term and its mass is proportional to $f_0^2$. The massless scalar ghost is interesting because it is the ghost found in most quantization schemes of pure Einstein gravity \cite{Alvarez-Gaume:16}. With some work, it can be shown to be harmless. The massive spin-two ghost arises uniquely from the $C^2$ term, and its mass is proportional to $\xi^2$.

In order to see what ultimately is a problem with ghost states, one can draw on the work on Lee-Wick models \cite{Lee:1969fy, Lee:1969fz, Lee:1970iw, Cutkosky:1969fq, Grinstein:2007mp,Grinstein:2008bg}. In these theories, states with negative norms are introduced much like the Pauli-Villars regulators, which combine with regular fields to produce a $q^{-4}$ fall-off of the combined propagators. While one might worry about unitarity and negative energies with the ghost states, it is not these features which are problematic. Due to the interactions of the theory, the ghost states are unstable and do not appear in the asymptotic spectrum\footnote{For other recent discussions regarding the emergence of real and complex ghosts in higher-derivative quantum gravity, see Refs.~\cite{Modesto:16,Accioly:17}.}. The theories can be shown to be unitary. However, what does occur is microscopic violations of causality due to the ghost states. While there are not gross large scale violations of causality, because the ghost states only propagate for a short time, the causal properties are uncertain on small scales by amounts of the order of the ghost width. This has been explicitly demonstrated in Refs.~\cite{Grinstein:2007mp, Grinstein:2008bg}. For example, by forming initial state localized wavepackets, the final arrival times will have portions of the wavepacket arriving earlier than usual expectations by amounts of order the width.

In our calculation, a similar effect in the propagator holds. Interactions shift the massive spin-two effect in the propagator away from the real axis, and it appears as a resonance rather than an asymptotic state. The sign of the propagator near the pole is ghost-like. However, the resulting amplitude near the pole will be explicitly shown to be unitary. We should also expect microscopic violations of causality as in the Lee-Wick models. To have this happen near the Planck scale in gravity does not seem outrageous. We would expect that the causal properties would become fuzzy in any quantum theory of gravity because the quantum fluctuations of whatever defines spacetime would lead to uncertainties in the causal structure. In our case these effects would be proportional to the Planck scale, although somewhat larger by an amount $1/\xi^4$ as the resonance width is narrow. We are willing to accept this feature as a property of our theory.

The other complaint about theories with extra derivatives in the fundamental Lagrangian is that they lead to a classical Hamiltonian which is not bounded from below, via the Ostrogradsky construction \cite{Woodard:2015zca}. It is not clear how relevant this classical result is. The path integral treatment based on the Lagrangian does not find the effect of such an instability. The Lee-Wick theories also seem not to be bothered by it. Our calculations do not show such an effect either. One notes that the classical Dirac Hamiltonian is also not bounded from below. The Dirac case is rendered stable by a modification of the quantization rules. There are at least three variations on modified quantization rules which lead to a well-behaved quantum Hamiltonian in the case of quartic derivative theories~\cite{Bender:2007wu, Salvio:2015gsi, Raidal:2016wop}. We do not take a position on these alternative rules. However, gauge theories and gravity are most simply quantized by the path integral formalism using the action, and we adopt these techniques. They seem to lead to a well-behaved quantum theory.

Finally, theories with quartic propagators at high energy violate the K\"{a}llen-Lehmann bound on the asymptotic behavior of propagators \cite{Kallen:1952zz, Lehmann:1954xi}. However, this bound does not apply for gauge fields - indeed it is violated in QCD \cite{Oehme:1979ai, Cornwall:2013zra}. Moreover, the microscopic violations of causality also violated the assumptions for the theorem. For both reasons, this bound is not relevant for the present theory.

Let us briefly discuss the quantization of the theory given by the action~(\ref{action1}) within the path integral treatment. Using the background-field method, one considers the parametrization
$g_{\mu\nu} = \bar{g}_{\mu\lambda}(e^h)^\lambda_{~\nu}$, where $\bar{g}_{\mu\nu}$ is a smooth background metric. One should add the necessary gauge-fixing terms concerning the Weyl part, $S_{\textrm{GF}}[h]$, and for the Yang-Mills term, $S_{\textrm{YM},\textrm{GF}}[A^{a}_{\mu},h]$. Moreover, one should also take into account the associated Faddeev-Popov ghost contributions $S_{\textrm{FP}}[\eta,\eta^{*},h]$ (for the Weyl part) and $S_{\textrm{gh}}[c,\bar{c},A^{a}_{\mu},h]$ (for the Yang-Mills part). In such contributions, $\eta$ is the ghost field related to the Weyl term and $c$ is the ghost field associated with the Yang-Mills term. For the Weyl contribution, one can consider the construction given in Refs.~\cite{bos,deBerredoPeixoto:2003pj}. On the other hand, the Yang-Mills part can be obtained from the usual Minkowski counterpart by the usage of general covariance and then one expands the metric as above. In this way, in the background-field method the partition function for our theory is given by
\bea
\hspace{-3mm}
Z[\bar{g}] &=&
(\det G^{\mu\nu})^{1/2}\int {\cal D}h {\cal D} \eta^{*} {\cal D}\eta {\cal D}A {\cal D}\bar{c} {\cal D} c
\nn\\
&\times&\exp \biggl\{i\Bigl\{S_{W}[h] + S_{\textrm{GF}}[h]
+ S_{\textrm{FP}}[\eta,\eta^{*},h] + S_{\textrm{YM}}[A^{a}_{\mu},h]
+ S_{\textrm{YM},\textrm{GF}}[A^{a}_{\mu},h]+ S_{\textrm{gh}}[c,\bar{c},A^{a}_{\mu},h]\Bigr\}\biggr\},
\eea
where $S_{W}[h]$ ($S_{\textrm{YM}}[A^{a}_{\mu},h]$) is obtained from~(\ref{action1}) by solely considering the Weyl (Yang-Mills) term expanded up to second order in $h_{\mu\nu}$, and $G^{\mu\nu}$ is the differential operator associated with the gravitational gauge-fixing term.

The one-loop divergences for the effective action evaluated from this theory have been calculated elsewhere, see for instance Refs.~\cite{bos,deBerredoPeixoto:2003pj,Julve:1978xn,Fradkin:1981hx,Fradkin:1981iu,Avramidi:1985ki,Avramidi:1986mj}. It is given by
\beq
\Gamma_{\textrm{div}} = -\frac{\mu^{d-4}\,\left(\alpha_1 + \alpha_2\right)}{d-4}
\int d^{d} x \sqrt{-\bar{g}}\,\bar{C}_{\mu\nu\alpha\beta}\bar{C}^{\mu\nu\alpha\beta}.
\eeq
The Weyl tensor $\bar{C}_{\mu\nu\alpha\beta}$ is built from the background metric $\bar{g}_{\mu\nu}$. We will conveniently specialize this result to a flat background in due course. In addition, in the above equation, $\alpha_1$ is the contribution coming from graviton loops, whereas $\alpha_2$ is the contribution coming from the non-abelian gauge field. Specifically:
\bea
\alpha_1 &=& \frac{199}{480 \pi^2} = \frac{N_q}{160 \pi^2}
\nn\\
\alpha_2 &=& \frac{D}{160 \pi^2}.
\eea
In order to include Standard Model particles, as well as interactions beyond the Standard Model, one would have to consider additional shifts in the $\alpha_2$ coefficient due to one-loop divergencies associated with such couplings. One finds
\beq
\alpha_2 \to \alpha'_2 = \frac{D + N_{SM}}{160 \pi^2} = \frac{N_{\infty}}{160 \pi^2},
\eeq
at very high energies.

\section{The spin-two propagator}

Here we focus on the study of the spin-$2$ propagator.  As discussed above, the free field propagating modes emerging in the model depend on the choice of gauge parameters as well as on the choice of field parametrization. However, the spin-$2$ part of the propagator of the $h_{\mu\nu}$ field maintains the same form irrespective of gauge parameters and field parametrizations~\cite{bos}. The most interesting feature of this exploration is the finite-width resonance in the propagator. This emerges from a competition of the $q^2$ and $q^4$ terms in the propagator. It appears with an opposite sign from usual expectation, and the imaginary part also changes sign from expectation, in a way that is consistent with the optical theorem. It is the remnant of a would-be ghost state, although because it is unstable it does not appear as a physical state in the asymptotic spectrum. Also of interest is the existence of three energy regions, in each of which the propagator has a distinct structure and/or different active degrees of freedom. In the highest energy region, the quadratic curvature terms are dominant, and this is reflected in the structure of the propagator.

In this section we set the cosmological constant to zero, and return to its influence in section \ref{general}. This allows us to expand around flat space. In this paper we will use the exponential parameterization of the fluctuation,  $g_{\mu\nu} = \eta_{\mu\lambda}(e^h)^\lambda_{~\nu} =\eta_{\mu\nu} + h_{\mu\nu}+ \frac12 h_{\mu\lambda}h^\lambda_{~\nu} +...$. One obtains the following free propagator (in momentum space)
\beq
iD^{0}_{\mu\nu\alpha\beta}(q) = -\frac{2i\xi^2}{(q^2 + i\epsilon)^2}\,{\cal P}^{(2)}_{\mu\nu\alpha\beta}.
\label{fey_prop}
\eeq
We will give an explicit expression for the spin-$2$ projector ${\cal P}^{(2)}_{\mu\nu\alpha\beta}$ below.

First let us work at low energies and include loop corrections. We treat all fields as massless - the small masses of Standard Model fields make no difference in the energy region of interest to us. In this case, the one-loop vacuum polarization can be written schematically as (temporarily suppressing the Lorentz indices), for $d \to 4$
$$
\Pi(q_{E}^2) = \xi^2 q_{E}^4\left(\frac{\mu}{q_{E}}\right)^{4-d} \frac{c_1}{d-4},
$$
where we considered dimensional regularization for regularizing the integrals and this result was obtained after a Wick rotation to the Euclidean space ($q_{E}$ is the Euclidean momentum). The scale factor $\mu$, with dimensions of mass, is inserted for dimensional reasons, and $c_1$ is a constant that does not contain poles as $d \to 4$. By rearranging the expression in terms of an exponential of a logarithm, one easily sees that, as $d \to 4$, the one-loop vacuum polarization will consist of a divergent part plus a finite part. In addition, the knowledge of the form of the divergence term gives us straightforwardly the logarithmic finite part, since they share the same coefficient\footnote{We keep only the logarithmic terms. Additional finite terms can be absorbed into the finite parts of the coupling constants and do not change our analysis.} .

All such discussions imply that the results presented at the end of the previous section allows us to determine explicitly the finite part of the one-loop vacuum polarization. By employing the following expansions of the different invariants up to second order in $h$ (in Minkowski background)
\bea
R^2 &=&
\partial_{\mu}\partial_{\nu}h^{\mu\nu} \partial_{\alpha}\partial_{\beta}h^{\alpha\beta}
- 2 \Box h \partial_{\mu}\partial_{\nu}h^{\mu\nu} + \Box h \Box h
\nn\\
R_{\mu\nu}^2 &=& \frac{1}{4}\Box h \Box h + \frac{1}{4}\Box h_{\mu\nu} \Box h^{\mu\nu}
+ \frac{1}{2}\partial_{\mu}\partial_{\nu}h^{\mu\nu} \partial_{\alpha}\partial_{\beta}h^{\alpha\beta}
-\frac{1}{2} \Box h \partial_{\mu}\partial_{\nu}h^{\mu\nu}
- \frac{1}{2} \partial_{\mu}\partial_{\nu}h^{\mu\alpha} \partial^{\beta}\partial^{\nu}h_{\beta\alpha}
\nn\\
R_{\mu\nu\alpha\beta}^2 &=&\Box h_{\mu\nu} \Box h^{\mu\nu}
-2 \partial_{\mu}\partial_{\nu}h^{\mu\alpha} \partial^{\beta}\partial^{\nu}h_{\beta\alpha}
+ \partial_{\mu}\partial_{\nu}h^{\mu\nu} \partial_{\alpha}\partial_{\beta}h^{\alpha\beta},
\eea
where some integrations by parts were carried out, one obtains the following one-loop contribution to the vacuum polarization
\bea
\Pi_{\mu\nu,\alpha\beta}(q^2) &=& -\frac{N^{(0)}_{\textrm{eff}}}{320\pi^2}
\ln\left(\frac{-q^2}{\mu^2}\right)
\left[\frac{1}{3} q_{\mu}q_{\nu}q_{\alpha}q_{\beta} + \frac{1}{2} q^4 I_{\mu\nu\alpha\beta}
+ \frac{1}{6} q^2(q_{\mu}q_{\nu}\eta_{\alpha\beta} + q_{\alpha}q_{\beta}\eta_{\mu\nu})
- \frac{1}{6} q^4 \eta_{\mu\nu}\eta_{\alpha\beta}
\right.
\nn\\
&-&\, \left.
\frac{1}{4} q^2 (q_{\mu} q_{\beta}\eta_{\nu\alpha} + q_{\mu} q_{\alpha}\eta_{\nu\beta}
+ q_{\nu} q_{\alpha}\eta_{\mu\beta} + q_{\nu} q_{\beta}\eta_{\mu\alpha})\right],
\eea
where $I_{\mu\nu\alpha\beta}$ is given by
\beq
I^{\mu\nu\alpha\beta} = \frac{1}{2}\left(\eta^{\mu\alpha}\eta^{\nu\beta}
+ \eta^{\mu\beta}\eta^{\nu\alpha}\right).
\label{identity}
\eeq
Here $N^{(0)}_{\textrm{eff}}$ is defined by the contributions of various matter fields to the one-loop divergence (within the Standard Model and beyond), normalized to that for gauge bosons. For the fields with spin $J\le 1$, this is
\beq
N^{(0)}_{\textrm{eff}} = N_V + \frac14 N_{1/2} + \frac16 N_S
\eeq
where $N_V,~ N_{1/2},~N_S$ are the number of gauge bosons, fermions and scalars respectively.  The contribution coming from graviton loops will be discussed in due course. The one-loop spin-$2$ propagator is then given by
\beq
i D_{\mu\nu\alpha\beta}(q^2) = i D^0_{\mu\nu\alpha\beta}(q^2)
+ iD^0_{\mu\nu\rho\tau}(q^2) [i\Pi^{\rho\tau,\gamma\delta}(q^2)] iD^0_{\gamma\delta\alpha\beta}(q^2).
\eeq
Now we rewrite $\Pi_{\mu\nu,\alpha\beta}(q^2)$ by introducing the set of the following projectors for symmetric second-rank tensors in momentum space~\cite{bos}
\bea
{\cal P}^{(2)}_{\mu\nu\rho\sigma} &=& \frac{1}{2}(\theta_{\mu\rho}\theta_{\nu\sigma}
+ \theta_{\mu\sigma}\theta_{\nu\rho}) - \frac{1}{3} \theta_{\mu\nu}\theta_{\rho\sigma}
\nn\\
{\cal P}^{(1)}_{\mu\nu\rho\sigma} &=& \frac{1}{2}(\theta_{\mu\rho}\omega_{\nu\sigma}
+ \omega_{\mu\rho}\theta_{\nu\sigma} + \theta_{\mu\sigma}\omega_{\nu\rho}
+ \omega_{\mu\sigma}\theta_{\nu\rho})
\nn\\
{\cal P}^{(0)}_{\mu\nu\rho\sigma} &=& \frac{1}{3} \theta_{\mu\nu}\theta_{\rho\sigma}
\nn\\
{\cal \bar{P}}^{(0)}_{\mu\nu\rho\sigma} &=& \omega_{\mu\nu}\omega_{\rho\sigma}
\eea
where
\bea
\theta_{\mu\nu} &=& \eta_{\mu\nu} - \frac{q_{\mu}q_{\nu}}{q^2}
\nn\\
\omega_{\mu\nu} &=& \frac{q_{\mu}q_{\nu}}{q^2}.
\eea
Such projectors do not form a complete basis in the corresponding space, and hence we must also add the following transfer operators
\bea
{\cal T}^{(0)}_{\mu\nu\rho\sigma} &=& \frac{1}{\sqrt{3}}\theta_{\mu\nu}\omega_{\rho\sigma}
\nn\\
{\cal \bar{T}}^{(0)}_{\mu\nu\rho\sigma} &=& \frac{1}{\sqrt{3}}\omega_{\mu\nu}\theta_{\rho\sigma}
\eea
in order to obtain a complete basis. In terms of such projectors the one-loop vacuum polarization can be written as
\bea
\Pi_{\mu\nu,\alpha\beta}(q^2) &=& -\frac{N^{(0)}_{\textrm{eff}}}{640\pi^2}
q^4 \ln\left(\frac{-q^2-i\epsilon}{\mu^2}\right)
{\cal P}^{(2)}_{\mu\nu\alpha\beta}.
\eea
The imaginary part of the vacuum polarization is found via the usual prescription
\beq
\ln (-q^2-i\epsilon) = \ln ( |q^2| ) -i\pi \theta(q^2)  \ \ .
\eeq

Using the usual orthogonality relations satisfied by the above projectors, one can iterate the vacuum polarization and find an intermediate form for the spin-$2$ propagator
\bea
D_{\mu\nu\alpha\beta}(q^2) &=& -\frac{\,{\cal P}^{(2)}_{\mu\nu\alpha\beta}}
{(q^2 + i\epsilon)^2\left[\frac{1}{2\xi^2(\mu)}+ \frac{ N^{(0)}_{\textrm{eff}}}{640\pi^2}
\ln\left(\frac{-q^2-i\epsilon}{\mu^2}\right)\right]}.
\label{graviton2}
\eea
The factor in square brackets defines the running coupling constant $\xi(q^2)$ at these energy scales (that is, below the resonance). That is, we define
\beq
\xi^2(q) = \frac{\xi^2(\mu)}{1 + \frac{\xi^2(\mu) N^{(0)}_{\textrm{eff}}}{320\pi^2}
\ln\left(\frac{|q^2|}{\mu^2}\right)}= \frac{320\pi^2}{N^{(0)}_{\textrm{eff}}
\ln\left(\frac{|q^2|}{\Lambda_{\xi}^2}\right)}.
\eeq
In addition, there are the effects of gravitational loops. At low energies these are treated using the effective field theory for gravity. As briefly discussed in the Introduction section, the Einstein-Hilbert gravitational action will be induced at low energies by the helper gauge interaction ~\cite{Adler:1980pg,Zee:1980sj,Zee:1981mk,Zee:1981ff,Adler:1982ri,Brown:1982am,Zee:1983mj}. In other words, an induced Einstein-Hilbert term is present in the effective action due to the scale-invariance breaking generated by loop effects. We have calculated this effect for QCD-like theories in a previous paper, finding a positive value for $\kappa^2$\cite{DM}. In addition to leading to a $q^2$ term in the propagator, the effects of loops of gravitons also provide logarithmic factors in the propagator. The loops carry the usual $i\epsilon$ convention.  Taking into account such modifications, one obtains that
\beq
\tilde{D}_{\mu\nu\alpha\beta}(q^2) = {\cal P}^{(2)}_{\mu\nu\alpha\beta} D(q^2),
\label{graviton-mod}
\eeq
where
\bea
D^{-1}(q^2) &=& \frac{q^2 + i\epsilon}{\tilde{\kappa}^2} - \frac{q^4}{2\xi^2(\mu)}
- \frac{N_{\textrm{eff}}}{640\pi^2} q^4\ln\left(\frac{-q^2-i\epsilon}{\mu^2}\right)
\eea
and $\tilde{\kappa}^2 = \kappa^2$ at the low energies which we are working presently. The pole at $q^2=0$ carries the usual $i\epsilon$ prescription because it was induced via the Yang-Mills interaction. Here $N_{\textrm{eff}}$ needs to include the graviton contribution, calculated within effective field theory~\cite{'tHV, Donoghue:1994dn}, such that
\bea\label{Neft}
N_{\textrm{eff}} &=& N_V + \frac14 N_{1/2} + \frac16 N_S + \frac{21}{6}
\nonumber \\
&=& \frac{21}{6} +N_{SM} + N_{BSM}
\eea
with $N_{SM}$ being the contribution from Standard Model particles and $N_{BSM}$ that of new physics beyond the Standard Model but below the scale of gravity. We neglect the latter in what follows. In the standard model with one Higgs doublet and three generations of fermions, one finds that $N_S = 2$,
$N_{1/2} = 45$, and $N_V = 12$, so that $N_{SM} = 283/12$.

The effect of graviton loops is somewhat different at high energies. Here the curvature-squared terms in the action are dominant, and the gravition progagator is quartic in the momentum. We treat this region by considering only the curvature-squared effect - the induced Einstein term is subdominant and is neglected. Here again the logarithmic effects are tied to one-loop divergences. However, the difference comes in that there is no imaginary part induced by these loops. This arises because the matrix element for the production of on-shell gravitons vanishes. This is known quite generally, in that the Weyl action produces no on-shell scattering amplitudes in flat space \cite{Dona:2015tra, Johansson:2017srf, Maldacena:2011mk, Adamo:2013tja, Adamo:2016ple}. In our situation, this can be seen directly.

The vertex involved is a triple graviton coupling between an off-shell graviton with invariant mass $q^2$ and two on-shell gravitons. The latter will be massless, transverse and traceless since the spectrum is defined by the induced Einstein action. Let us show that this triple graviton vertex vanishes. Consider the triple graviton vertex arising from the $R^2$ term in $R^2-\frac13 R_{\mu\nu}R^{\mu\nu}$. The curvature can be expanded around flat space in powers of the number of gravitational fields involved
\beq
R = R^{(1)} +R^{(2)} +R^{(3)}~+ ~...
\eeq
such that we can pull out the triple graviton term as
\beq\label{triple}
 \sqrt{-g}~R^2 = ...~+ \frac12 h^\lambda_\lambda R^{(1)} R^{(1)} + 2  R^{(1)} R^{(2)} ~+~...
\eeq
The on-shell condition corresponding to transverse, traceless fields at $q^2=0$ is that $R_{\mu\nu}=0$. This holds order by order in the expansion in the number of graviton fields. The off-shell graviton will be taken from one of the terms in Eq. \ref{triple}. However, no matter how this field is chosen, there always remains another curvature which satisfies $R^{(i)}=0$ leading to the vanishing of the vertex. This can be verified by direct computation. This argument also easily generalizes the $R_{\mu\nu}R^{\mu\nu}$.

The vanishing of this vertex implies that the logarithms do not pick up an imaginary part. This can be accomplished by using
$\frac12 \ln[(q^2)^2/\mu^4]$ instead of $\ln (-q^2-i\epsilon)/\mu^2$. This form can also be achieved by regularizing the propagator via
\beq
D\sim \frac{1}{q^4 +\epsilon^2} \ \ .
\eeq
This then implies that the final propagator involves two logarithmic factors
and takes the form quoted previously
\begin{eqnarray}
iD_{\mu\nu\alpha\beta} &=& i{\cal P}^{(2)}_{\mu\nu\alpha\beta} D_2(q) \nonumber \\
D^{-1}_2(q) &=& \frac{q^2 + i\epsilon}{\tilde{\kappa}^2(q)}- \frac{q^4}{2\xi^2(\mu)}- \frac{q^4 N_{\textrm{eff}}}{640\pi^2} \ln \left(\frac{-q^2-i\epsilon}{\mu^2}\right)- \frac{q^4 N_{q}}{1280\pi^2} \ln \left(\frac{(q^2)^2}{\mu^4}\right)  \ \ .
\end{eqnarray}
In the low energy region below the resonance, where the Einstein action dominates, we have $N_q=0$, and $N_{\textrm{eff}}$ is given by Eq. \ref{Neft}. Above the resonance but below the Planck scale, the Weyl action is dominant and we have $N_q = 199/3$ and $N_{\textrm{eff}}=N_{SM}$.
Finally above the Planck scale the gauge bosons of the helper gauge theory become active and we have $N_q=199/3$ and $N_{\textrm{eff}}=D+ N_{SM}=N_{\infty}$, where $D$ is the number of gauge bosons.

Now let us study the poles of the propagator given by Eq.~(\ref{graviton-mod}). We will show that the ghost state contained in the free propagator becomes unstable due to loop corrections. Clearly the spin-$2$ propagator~(\ref{graviton-mod}) has the standard pole at $q^2 = 0$, which means that the graviton remains massless at one-loop order. Note also that, for Euclidean $q^2_{E} = - q^2$ (or spacelike $q$), $D^{-1}(q^2_{E})$ has no other real zeros apart from $q_{E}^2 = 0$ for $q_{E}^2 > \Lambda_{\xi}^2$. In turn, for timelike $q$, $q^2 > 0$ and $D^{-1}$ picks up an imaginary part. Because we are in the weakly coupled region $\xi^2<< 1$, the pole will occur below the Planck scale.

Consider the propagator near the pole. If we expand using
\beq
q^2 = m_r^2 +\delta q^2
\eeq
and note that in this energy region $\tilde{\kappa}=\kappa$, we get the expansion of the inverse propagator
\bea
D^{-1}(q) &=& \frac{m_r^2}{\kappa^2} -m_r^4 \left[\frac{1}{2\xi^2(\mu)}+ \frac{N_{\textrm{eff}}}{640\pi^2}\ln \left(\frac{m_r^2}{\mu^2}\right)\right] \ \ \nonumber \\
&~&~ + \delta q^2
 \left(\frac{1}{\kappa^2}-2m_r^2\left[\frac{1}{2\xi^2(\mu)}+ \frac{N_{\textrm{eff}}}{640\pi^2}\left(\ln \left(\frac{m_r^2}{\mu^2}\right) +\frac12\right)\right] \right) \ \nonumber \\
 &~&~~ +i\pi m_r^4 \frac{N_{\textrm{eff}}}{640\pi^2}
\eea
Using
\beq
\frac{1}{2\xi^2(\mu)}+ \frac{N_{\textrm{eff}}}{640\pi^2}\ln \left(\frac{m_r^2}{\mu^2}\right) = \frac{1}{2\xi^2(m_r)}
\eeq
the location of the pole is the determined by the condition
\beq
m_r^2 = \frac{2 \xi^2(m_r)}{\kappa^2}
\eeq
and we have
\beq
D^{-1}(q) =- \frac{\delta q^2}{\kappa^2}
 \left(1+ \frac{N_{\textrm{eff}}\xi^2(m_r)}{320\pi^2} \right) +i\frac{2\xi^2m_r^2}{\kappa^2} \frac{N_{\textrm{eff}}}{640\pi}  \ \ .
\eeq
If we define the positive number $\gamma$ by
\beq
\gamma = {2\xi^2m_r^2} \frac{N_{\textrm{eff}}}{640\pi \left(1+ \frac{N_{\textrm{eff}}\xi^2(m_r)}{320\pi^2} \right)}
\eeq
and pull out an overall normalization (which can be absorbed in the normalization of the field), we have the propagator near the pole involving
\beq\label{proppole}
D(q) = \left[\frac{\kappa^2}{ 1+ \frac{N_{\textrm{eff}}\xi^2(m_r)}{320\pi^2} }\right]~\frac{1}{-\delta q^2 +i \gamma}= \left[ \frac{\kappa^2}{ 1+ \frac{N_{\textrm{eff}}\xi^2(m_r)}{320\pi^2}} \right]~\frac{-1}{\delta q^2 -i \gamma}  \ \ .
\eeq
This behavior is different from the usual structure for a propagator of an unstable particle. In the standard case we use
\beq
D(q) = \frac{1}{q^2 - (M-i \frac{\Gamma}{2})^2}
\eeq
and we would have near the pole, $m^2= M^2-\Gamma^2/4$, the behavior
\beq\label{standardpole}
D(q) =\frac{1}{\delta q^2 + i M\Gamma}  \ \ .
\eeq
We note that, aside from the overall normalization, the residue right on the pole is exactly the same as usual, with the correspondence
\beq
\frac{1}{i\gamma}= \frac{1}{iM\Gamma}  \ \ .
\eeq
However when the form of the propagators is defined by the sign of $\delta q^2$, this is composed of two unusual signs. Comparing the second version in Eq. \ref{proppole} with that of Eq. \ref{standardpole} we see that the overall normalization is the opposite of the expectation (i.e ghost-like), and also the imaginary part has the opposite sign from expectation. These two signs are connected in an important way. We can summarize these by a factor $Z$, such that
\beq
iD(q^2) \sim \frac{i Z}{q^2 - m_r^2 - iZ \gamma }
\eeq
where $Z = -1$.
As discussed in Refs.~\cite{Grinstein:2007mp,Grinstein:2008bg}, the overall minus sign in $Z$ is significant; such a sign is compensated by the unusual sign of the ghost propagator in such a way that the imaginary part of the forward scattering amplitude is positive, as it should be in order to obey the optical theorem. These signs also play a very important role in the unitarity of the scattering amplitude which we will discuss in the next section.

Our calculations show the emergence of three energy scales, defined by $\Lambda_{\xi}$, $\Lambda_{g}$, which is the gauge field scale mass, and $m_r$, the value for which the propagator presents a resonance, as discussed above. We take $\Lambda_{g} \gg \Lambda_{\xi}$. For energies below $\Lambda_{g}$, the gauge field becomes strongly coupled and confined. In addition, in this case one is justified in identifying the Planck scale with the Yang-Mills scale,  $\Lambda_{g} \equiv M_{p}$. For energies $\sim m_r$, the ghost particle goes as a resonance; since it is unstable, it will not appear in asymptotic states. Below such a scale one should take into account the contribution to the vacuum polarization coming from the gravitons in the effective-field-theory calculations. Below this energy scale the theory is satisfactorily described by the effective field theory approach.

In this analysis we have made an approximation of using only the dominant gravitational action in each region of interest. Specifically, at high energy where the Weyl action dominates, we have used only the quartic propagators and dropped the effect of the sub-dominant $q^2$ terms. In addition, we have considered only the one-loop corrections. It seems clear that a more complete treatment would use the full propagator self-consistently within the graviton contribution to the vacuum polarization. This would indeed be interesting to explore, and we hope to return to this calculation in the future.

\section{Unitarity of the scattering amplitude}

In this section we show, with explicit calculations, that the $J=2$ scattering amplitude in this theory is unitary at all the energy regions. This proceeds  through the graviton propagator in the s-channel and is the one which might be expected to be problematic due to the lowest order expectation for the ghost state. However, this feature turns out not to violate unitarity, and indeed it is the special properties of the resonance and the imaginary parts from loops which enforce unitarity. We proceed in three steps. First we describe the scattering of a single scalar field. This demonstrates how unitarity occurs in this channel and highlights the role of the various signs in the propagator and the imaginary parts. Then we generalize to multiple fields. Finally, there is a discussion of the graviton interactions at the highest energies due to the quadratic curvature terms in the action.

This calculation is based on previous work by Han and Willenbrock \cite{Han} and Aydemir {\it et al.} \cite{Aydemir}. As a first step, we work with a single real massless scalar field minimally  coupled to gravity and consider the reaction $\phi + \phi \to \phi + \phi$ through s-channel graviton exchange. We recall that it is through the s-channel that resonances and new unstable particles are usually probed. In addition, in the s-channel, the intermediate state satisfies $s > 0$, where $s$ is the Mandelstam variable describing the square of the total energy of the particles in the center-of-mass frame (invariant rest mass).

The Feynman rules associated with the interacting vertex for scattering amplitudes is to be extracted from the expression $(1/2) h_{\mu\nu} T^{\mu\nu}$. The s-channel amplitude for the process $\phi + \phi\to \phi+ \phi$ is given by
\beq
i {\cal M} = \left(\frac{1}{2} V_{\mu\nu}(q)\right)
\bigl[i{D}^{\mu\nu\alpha\beta}(q^2)\bigr]
\left(\frac{1}{2} V_{\alpha\beta}(-q)\right)
\eeq
where the energy-momentum of matter has the following on-shell matrix element (at the lowest order)
\beq
V_{\mu\nu}(q) = \langle p'| T_{\mu\nu} |p \rangle
= p_{\mu} p'_{\nu} + p'_{\mu} p_{\nu} - p \cdot p' \eta_{\mu\nu},
\eeq
with $p \cdot p' = q^2/2$, $p+p' = q$. Since we are only considering the effect of the scalar field, the logarithmic terms in the propagator only reflect the loop of that particle and are described by $N_q=0$; on the other hand, efective field theory calculations yield the value $N_{\textrm{eff}} =1/6$ for a single scalar field~\cite{Aydemir}. Employing the usual Mandelstam variables, one finds
\bea
{\cal M} &=& \frac{1}{8 s}\left(2tu - \frac{s^2}{3} \right) \bar{D}(s)
\nn\\
\bar{D}^{-1}(s) &=&\, \frac{1}{\tilde{\kappa}^2}\left\{1- \frac{\tilde{\kappa}^2 s}{2\xi^2(\mu)}
- \frac{\tilde{\kappa}^2 s N_{\textrm{eff}}}{640\pi^2} \ln \left(\frac{s}{\mu^2}\right)
+ \frac{i \tilde{\kappa}^2 s N_{\textrm{eff}}}{640\pi}\right\}
\label{amp}
\eea
where we used that $s + t + u = 0$. Now we perform a partial wave expansion with respect to the angular momentum $J$
\beq
{\cal M} = 16 \pi \sum_{J=0}^{\infty} (2J+1) T_{J}(s) P_{J}(\cos\theta)
\eeq
where $P_{J}(\cos\theta)$ are the Legendre polynomials that satisfy $P_{J}(1) = 1$.  Furthermore, we employ the parametrization $t = -s(1-\cos\theta)/2$ and $u = -s(1+\cos\theta)/2$. Hence using that $P_2(x) = (3x^2-1)/2$, one finds the following expression for the partial wave amplitude $T_2$:
\beq\label{t2amplitude}
T_2(s) = - \frac{N_{\textrm{eff}} s}{640 \pi}\,\bar{D}(s).
\eeq
In order to satisfy elastic unitarity, the scattering in the elastic channel must have $\textrm{Im} T_2 = |T_2|^2$. This is satisfied when the amplitude has the form
\beq
T_2(s) = \frac{A(s)}{f(s) -i A(s)}= \frac{A(s)[f(s)+i A(s)]}{f^2(s)+A^2(s)}
\eeq
for any real functions $f(s),~A(s)$. Since the imaginary part in the denominator comes from the logarithmic factor, the unitarity condition is a relation between the tree-level scattering amplitude which determines the $A(s)$ in the numerator and the logarithm in the vacuum polarization which determines the imaginary part in the denominator. For the elastic scattering of a single scalar, this relation is satisfied with
\beq
A(s) =  - \frac{N_{\textrm{eff}} s}{640 \pi}.
\eeq
We note that there is an important correlation between the unusual sign of the imaginary part in the propagator and the sign of the scattering amplitude which allows unitarity to be satisfied. It is however perhaps not that surprising that unitarity is obtained in this case because we are here simply expanding a unitary S matrix in perturbation theory.

If we now allow not only a single scalar but also other light fields plus gravitons in the theory, we are in principle faced with a multi-channel problem. An initial state of scalars can scatter into a final state of gravitons, and visa versa. The solution is given by Han and Willenbrock \cite{Han}, and consists of diagonalizing the scattering matrix. Having performed this diagonalization, the problem is back into that of a single channel elastic scattering. Let us first consider this in the effective field theory limit, below the resonance where graviton scattering is determined by the Einstein action only (i.e. $N_q=0$). The diagonalization of Han and Willenbrock can be extended to include the graviton channel, and the result is identical to that of Eq.~(\ref{t2amplitude}) except with a generalized value of $N_{\textrm{eff}}$, given by Eq.~(\ref{Neft}). So we see that unitarity is again satisfied.

Finally, we turn to the high energy region where the graviton propagator arises dominantly  from the terms quadratic in the curvature. Consistent with the approximation given above, we here drop consideration of the Einstein term in the gravitational action and only consider the quadratic terms. As we described, these give no coupling for on-shell gravitons. Likewise in the propagator, the imaginary parts only arise from the matter particles and not the gravitons. The scattering amplitude then takes the form
\beq\label{t2amplitudehigh}
T_2(s) = - \frac{N_{\textrm{eff}} s}{640 \pi}\left\{- \frac{s}{2\xi^2(\mu)}- \frac{s N_{\textrm{eff}}}{640\pi^2} \left[\ln \left(\frac{s}{\mu^2}\right)-i\pi\right]- \frac{s N_{q}}{1280\pi^2} \ln \left(\frac{s^2}{\mu^4}\right)
\right\}^{-1}.
\eeq
In this region, we have
\beq
N_{\textrm{eff}} = N_{\infty} = D + N_{SM}~~~~~~~N_q = \frac{199}{3}.
\eeq
Again, unitarity is satisfied. The consistent decoupling of the on-shell graviton states in both the numerator and denominator was important in this regard.

We repeat the comment from the previous section that an improved treatment would include the full propagator self-consistently in loops, as we hope to do in the near future. This will be especially instructive in the unitarity calculation as it has the potential to couple in on-shell gravitons even at the highest energies. However it appears from the general construction of the elastic channel that we would expect unitarity to be satisfied in such a treatment also.

\section{More general results}\label{general}

The discussion above has focussed on what we feel is the most important and novel aspects of the present theory. There are two main topics connected to other terms in the action, 1) the cosmological constant and 2) the $R^2$ term in the quadratic action which is associated with the scalar degree of freedom in quadratic gravity.

Let us start with the latter effect. Even if excluded from the initial action, the $R^2$ interaction will be induced by the helper gauge interaction, much like the Einstein action is induced. This effect was first described by Brown and Zee \cite{Brown:1982am}. We discuss this in our Appendix below and calculate it within QCD. The result, in the notation of Eq. \ref{quadraticorder}, is positive and has the value
\begin{equation}\label{f0}
  \frac{1}{6f_0^2}=   0.00079 \pm 0.00030.
\end{equation}
Because this is dimensionless, it would hold in a scaled-up version of QCD. However, the result has an unusual feature that this value is only to be applied when working below the Planck energy. It can be seen from the derivation of the Adler-Brown-Zee formalism that the calculation is accomplished by Taylor expanding the gravitational field on longer wavelengths than the Planck scale arising from the helper gauge interaction. The induced effect vanishes at energies above the Planck scale, much like the induced value of the Newton constant also vanishes there.

It is also possible to have a ``bare" value of $f_0$ in the original action. In this case the original symmetry is scale invariance rather than local conformal invariance\footnote{Indeed calculations exist that show that divergences in $f_0 $ is generated perturbatively at higher loop order even if one starts from a conformally invariant initial action. See the discussion of Salvio and Strumia \cite{Strumia}. It would be interesting to understand if there were regularization schemes which could prevent such divergences.} . In this case there is a scalar sector to analyze. The interesting feature here is that the massive scalar that appears is not a ghost state. It has the usual positive residue, and is not problemmatic. Indeed, the scalar propagator has been calculated elsewhere and it is given by~\cite{Strumia,Alvarez-Gaume:16}
\beq
D^{(0)}_{\mu\nu\alpha\beta}(q^2) =
\left(\frac{q^4}{f_0^{2}} - \frac{2q^2}{\tilde{\kappa}^2}\right)^{-1} {\cal P}^{(0)}_{\mu\nu\alpha\beta}
= \frac{\tilde{\kappa}^2}{2}\left(\frac{1}{q^2 - M_0^{2}}- \frac{1}{q^2}\right)
{\cal P}^{(0)}_{\mu\nu\alpha\beta}
\label{prop-scalar}
\eeq
where $M_0^{2} = 2f_0^{2}/\tilde{\kappa}^2$. One can easily identify the massive scalar mode, as well as a (massless) ghost state. However, with some effort one can show that the latter is the standard ghost that emerges also in General Relativity, so it is a harmless non-propagating state which can be removed within a gauge transformation~\cite{Alvarez-Gaume:16}. In turn, the first term describes the massive scalar excitation, which is in fact a propagating mode, as alluded to above.  In this case one can also define a running coupling constant $f_0^2(q)$ through the renormalization group equation~\cite{Strumia}:
\beq
\mu\frac{d f_0^2}{d\mu}  = \frac{1}{16\pi^2}\left(\frac{5 \xi^4}{3} +  5 \xi^2 f_0^2 + \frac{5}{6} f_0^4\right).
\eeq
Observe that the coupling $f_0$ is not asymptotically free, unless $f_0^2 < 0$, which would lead to a tachionic instability $M_0^2 < 0$.

It also would not be a problem for this theory if we were to give up the initial scale/conformal invariance, as long as the energy scale of the intrinsic $G^{-1}$ and cosmological constant were small compared to the Planck scale determined by the helper gauge interaction. For example for weakly coupled Weyl gravity, we have seen that the intrinsic scale in the running coupling is $\Lambda_{\xi}$, which can be $10^{-1006}$ eV or even much lower. If the intrinsic Newton's constant and cosmological constant were govern by that scale, the phenomenology of this model would be essentially unchanged.

Finally, there is the problem of the induced cosmological constant. In a QCD-like theory, this is given by
\begin{equation}
\Lambda_{\textrm{ind}}= \frac14 \langle 0 |T^\mu_{~\mu}| 0 \rangle
= \frac14 \left\langle 0 \bigg|\frac{\beta(g)}{2g} F^a_{\mu\nu}F^{a\mu\nu}\bigg| 0 \right\rangle \ \ .
\end{equation}
For the standard value of the gluonic condensate, this yields $\Lambda_{\textrm{ind}} = - 0.0034$~GeV$^4$. Scaling a QCD-like theory up to the Planck mass would yield a very large negative value of the cosmological constant, of order the Planck scale.

In the presence of this cosmological constant, we need to expand about anti-de Sitter space rather than about flat space. We can show that the AdS solution is unchanged by the presence of quadratic terms.  Using the exponential
parametrization around a background field
\beq
g_{\mu\nu} = \bar{g}_{\mu\lambda}(e^h)^\lambda_{~\nu}
\eeq
and discarding total derivatives, we find the expansion of $\sqrt{-g}({\cal L}_{\rm EH}+{\cal L}_{q})$ where
\bea\label{quadnormal}
\sqrt{-g}{\cal L}_{\rm EH} &=&\sqrt{-\bar{g}} \left[ -\Lambda + \frac{2}{\kappa^2}\bar{R} \right. \ \ \nonumber \\
&~&~~+\frac{1}{\kappa^2}\left(\bar{g}^{\mu\nu}\bar{R}-2\bar{R}^{\mu\nu}-\frac{\Lambda\kappa^2}{2}\bar{g}^{\mu\nu}\right) h_{\mu\nu} \ \nonumber \\
&~&~~-\frac{\Lambda}{8} (h^\lambda_\lambda)^2 + \frac{\bar{R}}{4\kappa^2}(h^\lambda_\lambda)^2 -\bar{R}^{\mu\nu}\left( h_{\mu\nu}h^\lambda_\lambda - h_{\mu\beta}h^\beta_\nu \right) \ \nonumber \\
&~&~~\left.+\frac{1}{\kappa^2}\left(h^\lambda_{\lambda,\nu}h^{\nu,\beta}_\beta -h^\beta_{\alpha,\nu}h^{\nu,\alpha}_\beta +\frac12h^\beta_{\alpha,\nu}h^{\alpha,\nu}_\beta -\frac12 h^\lambda_{\lambda,\nu}h^{\beta,\nu}_\beta\right)\right]  \ \ .
\eea
For the quadratic terms, ${\cal L}_q$, we use $R = \bar{R} + R^{(1)}+...$, where $R^{(1)}$ is linear in the quantum fluctuation
\beq\label{Rexp}
R^{(1)} = h^{\lambda~~,\nu}_{\lambda,\nu} -h_{\mu\nu}^{~~~,\nu,\mu} - \bar{R}_\nu^\mu h_\mu^\nu \ \,
\eeq
resulting in
\beq\label{quadexp}
\sqrt{-g} R^2 = \sqrt{-\bar{g}}\left[\bar{R}^2 +\left(2 \bar{R}R^{(1)} + \frac12 \bar{R}^2h^\lambda_\lambda\right)+ \left((R^{(1)})^2 +\bar{R}R^{(1)} h^\lambda_\lambda +\frac18 (h^\lambda_\lambda)^2 \bar{R}^2\right)+...\right]  \ \ .
\eeq
However, for constant curvature spacetimes with $\bar{R}_{\mu\nu} = \beta g_{\mu\nu}, ~ \bar{R}=4\beta$, with $\beta$ being a constant, the linear term $\left(2 \bar{R}R^{(1)} + \frac12 \bar{R}^2h^\lambda_\lambda\right)$ vanishes aside from a total derivative. Moreover, for such a spacetime the background Weyl tensor vanishes $\bar{C}_{\mu\nu\alpha\beta}=0$. While in general $\sqrt{-g} C^2$ would have a similar expansion to Eq. \ref{quadexp}, in this spacetime there will be no linear term in the expansion because of the vanishing of the background Weyl tensor. Therefore the lowest order Einstein equation remains unchanged and we recover the AdS curvature $R_{\mu\nu} =\frac14 \kappa^2\Lambda \bar{g}_{\mu\nu}$ and the background dependent terms of Eq. \ref{quadnormal} reduce to
\beq\label{explicit}
-\frac{\Lambda}{8}\left[  (h^\lambda_\lambda)^2 -2h^\beta_\alpha h^\alpha_\beta \right]  \ \ .
\eeq
The expansion of ${\cal L}_q $ is then
\bea
\sqrt{-g} {\cal L}_q = \sqrt{-\bar{g}} \left[\frac{1}{6 f_0^2}\bar{R}^2+\frac{1}{6 f_0^2} \left((R^{(1)})^2 +\bar{R}R^{(1)} h^\lambda_\lambda +\frac18 (h^\lambda_\lambda)^2 \bar{R}^2\right)-\frac{1}{2\xi^2}C^{(1)}_{\mu\nu\alpha\beta}C^{(1)\mu\nu\alpha\beta}\right]   \ \ .
\eea
The bilinear terms in the background field expansion of ${\cal L}_q $ can be recovered by use of the identity of Eq. \ref{weylidentity} and
\beq\label{Ricciexp}
R^{(1)}_{\mu\nu} = \frac12\left[h^\lambda_{\lambda,\mu,\nu} -h^\lambda_{\mu,\nu,\lambda} -h^\lambda_{\nu,\mu,\lambda}+h^{~~~,\lambda}_{\mu\nu~~~,\lambda} \right]
\eeq
as well as Eq. \ref{Rexp}.

While probably all quantum theories of gravity suffer from having a Planck-scale cosmological constant, it is particularly embarrassing for a theory which starts from a scale invariant initial action. In such a case, one cannot add a ``bare'' cosmological constant in order to cancel the induced one. We need to remove or severely suppress the induced cosmological constant. We present a set of possible solutions without having a perfectly compelling choice.

Holdom \cite{Holdom:2007gg} has raised the question of whether the vacuum energy in massless QCD could in fact vanish. This is in part because the present lattice estimates are uncertain enough to include a vanishing value. The calculations are made more difficult by the presence of a hard dimensional cutoff - the lattice spacing - which adds a dimensionful ingredient to the theory beyond just the running coupling. It would be an important step for induced gravity theories if this question could be definitively answered.

One of the conditions under which the vacuum energy would certainly vanish in a Yang-Mills theory is that the beta function could vanish at zero energy. This can happen if there is an infrared fixed point in the beta function at a finite coupling - the Caswell-Banks-Zaks fixed point \cite{Caswell:1974gg, Banks:1981nn}. This can emerge from a competition of terms of different order in the beta function, i.e. $\beta(g)= -bg^3 + cg^5$ with $b,~c>0$. However, the phenomenology of the IR phase is less-well understood. There is still a dimensional parameter in the running coupling at higher energies, and the perturbative contribution to the induced Einstein action would still exist. However it is not clear if a positive Newton's constant would result. This option deserves further study.

One of the present authors has proposed that the spin connection, which appears naturally as a gauge field in gravitational theories, could form an asymptotically free gauge theory if treated as an independent field \cite{Donoghue}. The Euclidean version of such a theory would involve the symmetry $O(4)$ and would be confined. This involves six field strengths, $F^{[a,b]}_{\mu\nu}$ where $[a,b]=a,b -b,a$ is the antisymmetric combination of Lorentz indices $a,~b = 0,1,2,3$. By symmetry, each of the fields contributes equally to the Euclidean vacuum value of $F^2$. When continued back to Lorentzian space-time, the symmetry becomes $SO(3,1)$. In the vacuum energy, the products $F^{[01]}F_{[01]}, ~F^{[02]}F_{[02]},~ F^{[03]}F_{[03]}$ change sign due to the Minkowski metric, while the spacelike ones $F^{[12]}F_{[12]}, ~F^{[13]}F_{[13]}, ~F^{[23]}F_{[23]}$ carry the same sign. The Lorentzian symmetry then forces the vacuum expectation value of the trace anomaly to vanish. If this non-compact group makes sense as the helper gauge interaction, then it would not suffer from the large cosmological constant problem.

We have been exploring this theory in the limit where the Weyl coupling constant $\xi$ stays small, or more precisely that the scale involved in the running coupling $\xi(q)$ is very much smaller than the Planck scale. This is chosen mainly for our convenience in analysing the theory. However, if the scale in the gravitational running constant is of the same order as the Planck scale, it will also become strongly coupled. In this case, it would induce extra contributions to the induced Newton constant and to the cosmological constant. Almost nothing is known about this situation, although Adler \cite{Adler:1982ri} has developed some of the relevant formalism. In the absence of knowledge, one can engage in wishful thinking that perhaps a mechanism can be found such that the gravitational contribution to the cosmological constant cancels the gauge contribution.

Another approach could be to have the helper gauge theory be supersymmetric. This would lead to a vanishing of the cosmological constant if the supersymmetry is still exact at the Planck scale. One would need supersymmetry breaking at a lower scale, and potentially a much smaller value of $\Lambda$.

Finally, one can have a version of the theory in which the energy density of the vacuum does not gravitate. This is accomplished by fixing the determinant of the metric to a constant, such that the vacuum energy term in the action does not involve the gravitational field. We discuss this unimodular option in the next section.

\section{Unimodular GAQ gravity}

Because of the difficulties associated with the cosmological constant, it is worth considering a unimodular version of the theory. In unimodular general relativity \cite{Anderson:1971pn, vanderBij:1981ym, Buchmuller:1988wx, Unruh:1988in, Weinberg:1988cp, Henneaux:1989zc, Percacci:2017fsy}, the determinant of the metric is set equal to a constant, which can be chosen to be unity,
\beq
\sqrt{-g} =1
\eeq
and the symmetry of the theory is reduced to volume-preserving diffeomorphisms which preserve this condition. The initial equations of motion differ from those of general relativity, because the $\sqrt{-g}$ factor is not varied, but there is an additional constraint on the solutions associated with energy conservation. When this constraint is imposed, the equations are exactly those of general relativity, including the possibility of a cosmological constant term which emerges as an integration constant of the constraint equation. Percacci \cite{Percacci:2017fsy} has a recent treatment of unimodular theories which makes it clear that the equivalence of the unimodular theory to the general one is not just a property of the Einstein action, but would also be true for a theory quadratic in the curvature, such as we are considering here.

However, the striking advantage of a unimodular theory is the different status of the cosmological constant. In usual metric theories, the cosmological constant is tied to the energy density of the vacuum, and appears as a constant term in the Lagrangian. The many large contributions to this energy density form part of the ``cosmological constant problem''. In unimodular theories, this energy density in the Lagrangian (which we have called $\Lambda$ above) does not couple to gravity and is irrelevant. The cosmological constant which appears in the equations of motion arises from the initial condition of the constraint equation. We can call this constraint parameter $\Lambda_u$. Of course, we do not yet have a good theory of this initial condition. For our purposes $\Lambda_u$ can have any magnitude and either sign. Unimodular theories solve this part of the cosmological constant problem by decoupling the constant that appears in Einstein's equation $\Lambda_u$ from the vacuum energy $\Lambda$. In the theory considered in this paper, the unimodular constraint would remove the worry about the large energy density found when one scales up QCD-like theories as the helper gauge theory.

Our starting point can be close to the unimodular case. Consider first the theory with a gauge interaction and the Weyl-squared action, as in Eq. 1. This has local conformal symmetry, and it has one less propagating degree of freedom compared to general quadratic gravity. In the exponential parameterization, this is the scalar trace $h=h^\lambda_{~\lambda}$. This ``conformal mode'' is missing from the action to all orders in pure 4D, not just in the expansion about flat space.

This set-up differs from a unimodular version in the path integral measure. In a general theory, the conformal mode is included in the path integration, whereas in a unimodular one it is not included. When one regulates the general theory, this mode can in principle get fed back into the action. If one regulates dimensionally, the conformal mode actually appears in the action when the dimension D is not equal to 4. If one regulates with a cutoff, then the mode is presumably sensitive to the cut-off when done in a gauge invariant manner. So the path integral measure is one way that the general action action differs from a unimodular one.

Let us then remove the conformal mode from the path integration. The background field expansion for the final theory changes. From the analysis shown above using the exponential parameterization, we must set $h= h^\lambda_{~\lambda}=0$. The $R^2$ action could still be induced, or even appear initially.
The remaining background field expansion has a slightly modified structure in the unimodular case. Generally, when we expand to second order in the fluctuation, we take the first order term to vanish by the equations of motion. But for unimodular gravity the first order term is not equivalent to the final equations of motion. In particular it does not include the constraint that brings in the parameter that plays the role of the cosmological constant. In particular, the first order term in the expansion is
\begin{equation}
\hat{h}_{\mu\nu}(R^{\mu\nu}-\frac14 \bar{g}^{\mu\nu}R)
\end{equation}
where the fluctuating field with the conformal mode removed is
\beq
\hat{h}_{\mu\nu}={h}_{\mu\nu}-\frac14 \bar{g}^{\mu\nu}h^{\lambda}_{~\lambda}  \ \  .
\eeq
This is the equation of motion before the constraint is imposed. However, both this original equations of motion and the constraint are separately satisfied. This implies that the first order variation vanishes as usual. When we reach second order in the fluctuation there are also background curvatures here. Here we should use the background curvatures which satisfy the final equations of motion. This leads to two differences in the final result. One is that the fluctuating field is traceless in the exponential representation. This removes the vacuum energy $\Lambda$ from the quadratic action. The other is that the equations of motion will involve the constraint-generated cosmological constant rather than the one connected with the vacuum energy. That is, one uses $R_{\mu\nu}= \frac14 \kappa^2 \Lambda_u g_{\mu\nu}$. For example, these changes yield the leading low energy action for AdS or or dS spaces as
\begin{equation}\label{quaduni}
S_{EH}= \int d^4x \frac{1}{\kappa^2} \left[-\hat{h}^\beta_{\alpha,\nu}\hat{h}^{\nu,\alpha}_\beta +\frac12 \hat{h}^\beta_{\alpha,\nu}\hat{h}^{\alpha,\nu} +\frac{\Lambda_u}{4}\hat{h}^\mu_\nu \hat{h}^\nu_\mu \right]
\end{equation}
instead of Eqs. \ref{quadnormal}, \ref{explicit}.

\section{Summary}

There are several features of this theory which differ from usual expectation.  We list these here:

\begin{itemize}

\item Although the initial theory was scale invariant, the Planck scale and the Einstein term in the action were induced by the helper gauge theory interaction via dimensional transmutation. This allows the low energy spectrum of gravitons to be the usual one.

\item The spin-two graviton propagator has a resonance at a scale $\xi M_P$. This is the remnant of the ghost state, but is unstable and does not exist as an asymptotic state in the spectrum. Related work on Lee-Wick models indicates that we should expect microscopic violations of causality on time scales given by the inverse of the width of the resonance, which is $1/\xi^2 M_P$.

\item The propagator sign at the resonance is opposite usual expectation, and likewise the imaginary part has the opposite sign from usual expectation. These two features combine to give the residue at the resonance the proper sign, consistent with the optical theorem.

\item There are three kinematic regions with different physical content. At low energies, below the resonance, the usual effective field theory description is valid. In the intermediate energies between the resonance and the Planck scale, the quadratic curvature terms dominate for the gravitons, but the helper gauge interaction is not yet active. At high energies, the theory can be asymptotically free with both the gauge and graviton interactions active.

\item When treated using only the interactions from the quadratic curvatures at ultra-high energies, the propagator does not pick up imaginary parts from the graviton loops, because the coupling to on-shell states vanishes. 
    
\item The spin-two partial wave amplitude - which is the dangerous one due to the original ghost pole - has been shown to be unitary in all energy regions. This is non-trivial and is tied to the special properties of the propagator.

\item The gravitational interaction remains weakly coupled at all energies. The usual growth of amplitudes with increasing energy is tamed by the transition to quartic propagators at high energy.

\item The $R^2$ term in the action is also induced by the helper gauge interaction even if it was not present in the initial theory. We calculate its coefficient in QCD-like theories. However, this has the feature that the induced coupling is only present below the Planck scale, and disappears above it.

\end{itemize}

The role of the helper gauge interaction in the theory is primarily to dynamically generate the Planck scale, while still keeping the gravitational interaction weakly coupled. This is the feature which allows the theory to be under reasonable theoretical control. We understand the dynamics of confined gauge theories from decades of study of QCD and related theories. By keeping the gravitational interaction weakly coupled, we can use perturbation theory to describe it and to bypass a strongly coupled theory of gravity. One can imagine other limits of this same theory in which the gravitational interaction is also strongly interacting. Or one could consider other methods for inducing the Planck scale, perhaps with scalar fields. However, the variation considered here, with weakly coupled gravity and a strongly interacting gauge theory, seems to be a very good laboratory for UV complete quantum field theories of gravity.

The theory has passed the tests that we have explored thus far. The naive fears concerning ghosts and unitarity violation have not been realized. The mechanisms for avoiding these have been non-trivial, and required the signs and amplitudes to work out in a coordinated manner. Further work is needed in order to study amplitudes with gravitational loops in this theory, and we hope to turn to this in the future. In addition, the cosmological constant problem remains, although we have suggested possible solutions, including a unimodular version of the theory. Overall, the possibility of using this theory as a QFT-based ultraviolet completion of quantum gravity remains promising.

\section*{Acknowledgements} We thank Alberto Salvio, Alessandro Strumia, Roberto Percacci, Guy Moore, David Kastor, Larry Ford, J.J. Carrasco and David Kosower for useful discussions. This work has been supported in part by the National Science Foundation under grant NSF PHY15-20292.

\section*{Appendix - The calculation of the induced $R^2$ term in the action. }

Here we derive the induced $R^2$ term within the Adler-Zee-Brown approach. The gravitational effective action reads
\beq
e^{iS_{\textrm{eff}}[g_{\mu\nu}]} = \int d\phi e^{i S[\phi,g_{\mu\nu}]},
\eeq
where $\phi$ represents generically the matter fields and $S[\phi,g_{\mu\nu}]$ describes matter fields on a curved background.  As discussed above, we are considering the Yang-Mills field as the matter field responsible for inducing the effective gravitational action. Since $S_{\textrm{eff}}[g_{\mu\nu}]$ is a scalar under general-coordinate transformations, it may be represented as the integral over the manifold of a scalar density, which for slowly varying metrics can be formally developed in a series expansion in powers of $\partial_{\lambda}g_{\mu\nu}$
\bea
S_{\textrm{eff}}[g_{\mu\nu}] &=& \int d^{4}x \sqrt{-g} {\cal L}_{\textrm{eff}}[g_{\mu\nu}]
\nn\\
{\cal L}_{\textrm{eff}}[g_{\mu\nu}] &=& {\cal L}^{(0)}_{\textrm{eff}}[g_{\mu\nu}]
+ {\cal L}^{(2)}_{\textrm{eff}}[g_{\mu\nu}]
+ {\cal L}^{(4)}_{\textrm{eff}}[g_{\mu\nu}]
+ {\cal O}[(\partial_{\lambda}g_{\mu\nu})^{6}]
\nn\\
{\cal L}^{(0)}_{\textrm{eff}}[g_{\mu\nu}] &=& \Lambda_{\textrm{ind}},\,\,\,
{\cal L}^{(2)}_{\textrm{eff}}[g_{\mu\nu}] = \frac{R}{16\pi G_{\textrm{ind}}},\,\,\,
{\cal L}^{(2)}_{\textrm{eff}}[g_{\mu\nu}] = \frac{1}{6 f^2_\textrm{ind}} R^2.
\eea
Let us derive a representation of the induced $f^2$ in terms of the vacuum expectation value of products of the stress-energy tensor $T_{\mu\nu}$ of the matter fields. For that, let us study the response of the action functional to an external classical gravitational field $g_{\mu\nu} = \eta_{\mu\nu} + h_{\mu\nu}$ treating $h_{\mu\nu}$ as a small perturbation. The response of quantum fields to an arbitrary external field is described by the generating functional of connected Green's functions:
\bea\label{Wexp}
iW[h] &=& -\frac{i}{2} \int d^{4}x h_{\mu\nu}(x) \langle T^{\mu\nu}(x) \rangle
+ \frac{i}{4} \int d^{4}x h_{\mu\nu}(x)h_{\alpha\beta}(x) \langle \tau^{\mu\nu\alpha\beta}(x) \rangle
\nn\\
&+& \frac{i^{2}}{2!}\left(\frac{1}{2}\right)^{2} \int d^{4}x \int d^{4}y\,
h_{\mu\nu}(x)h_{\rho\sigma}(y) \langle T \{{\bar T}^{\mu\nu}(x){\bar T}^{\rho\sigma}(y)\}\rangle + \ldots...
\eea
where $\langle \ldots \rangle = \langle 0| \ldots |0 \rangle$ denotes vacuum expectiation value, $T^{\mu\nu}$ is the energy-momentum tensor of the Yang-Mills field, given by
\beq
T_{\mu\nu} = - F^{a}_{\lambda\mu}F^{a\lambda}_{\ \ \nu}
+ \frac{1}{4}\eta_{\mu\nu} F^{a}_{\alpha\beta}F^{a\,\alpha\beta}.
\eeq
and ${\bar T}^{\mu\nu}(x) = T^{\mu\nu}(x) - \langle T^{\mu\nu}(x) \rangle$.  The second term on the right-hand side of Eq. \ref{Wexp} is needed for consistency; it plays a role in the cosmological constant sum rule~\cite{DM}. We consider, for arithmetical simplicity, that $h_{\mu\nu} = (1/4)\eta_{\mu\nu}h$. We choose $h$ to be a slowly varying over the scale of the gauge interaction so that one can expand
\beq
h(y+z) = h(y) + z^{\mu}\partial_{\mu}h(y) + \frac{1}{2} z^{\mu} z^{\nu} \partial_{\mu} \partial_{\nu}h(y) + \ldots
\eeq
In this way, with $W = \int d^4 x \sqrt{-g} {\cal L}_{\textrm{eff}}$, one finds
\bea
i\sqrt{-g} {\cal L}_{\textrm{eff}}[g_{\mu\nu}] &\approx&
-\frac{i}{2}  h(x) \left[\frac{1}{4}\langle T(x) \rangle\right]
+ \frac{i}{16}(h(x))^{2}\left[ \frac{1}{4} \langle \tau^{\mu}_{\ \mu}{}^{\alpha}_{\ \alpha}(x) \rangle \right]
- \frac{1}{16} (h(x))^{2}\int d^{4} z \left[\frac{1}{8}K(z)\right]
\nn\\
&+& \frac{1}{2^{10}}(\partial_{\mu} h)^{2}\int d^{4}z\,z^2\left[K(z)\right]
- \frac{1}{24576}(\partial_{\mu}^2 h)^{2}\int d^{4}z\,(z^2)^2\left[K(z)\right]
\eea
where integrations by parts in the two last terms were performed, $K^{\mu\nu\rho\sigma}(x-y) = \langle T \{{\bar T}^{\mu\nu}(x){\bar T}^{\rho\sigma}(y)\}\rangle$, $T = \eta_{\mu\nu} T^{\mu\nu}$ and $K = \eta_{\mu\nu}\eta_{\rho\sigma} K^{\mu\nu\rho\sigma}$. In turn
\bea
\sqrt{-g} R &\approx& \frac{1}{4}\left[\partial_{\alpha}h_{\lambda\kappa}\partial^{\alpha}
\left(h^{\lambda\kappa} - \frac{1}{2}\eta{\lambda\kappa} h\right)
- 2 \partial^{\mu}\left(h_{\lambda\mu} - \frac{1}{2}\eta_{\lambda\mu}h\right)
\partial_{\kappa}\left(h^{\lambda\kappa} - \frac{1}{2}\eta^{\lambda\kappa} h\right)\right]
= - \frac{3}{32}(\partial_{\mu} h)^{2},
\nn\\
\sqrt{-g} R^2 &\approx&
\partial_{\mu}\partial_{\nu}h^{\mu\nu} \partial_{\alpha}\partial_{\beta}h^{\alpha\beta}
- 2 \Box h \partial_{\mu}\partial_{\nu}h^{\mu\nu} + \Box h \Box h
= \frac{9}{16} (\partial_{\mu}^2 h)^{2}
\eea
and also $\sqrt{-g} = 1 + (1/2)h + (1/16)h^2$. Equipped with such results one is able to derive representations for the induced cosmological constant and induced Newton's constant as described in Refs.~\cite{Adler:1980pg,Zee:1980sj,Zee:1981mk,Zee:1981ff,Adler:1982ri,Brown:1982am,Zee:1983mj} and recently investigated in Ref.~\cite{DM} for the gluonic QCD case. In the present context, we need an expression for the induced coupling constant appearing before the $R^2$ term. One finds
\beq
\frac{1}{6 f^2_\textrm{ind}} = \frac{i}{13824}\int d^{4}z\,(z^2)^2 \langle T \{{\bar T}(z){\bar T}(0)\}\rangle.
\label{induced}
\eeq
Working in Euclidean space, one obtains
\beq
\frac{1}{6 f^2_\textrm{ind}} = \frac{1}{13824}\int d^{4}x\,(x^2)^2 \langle T \{{\bar T}(x){\bar T}(0)\}\rangle.
\label{induced2}
\eeq
After performing a change of variables $x^{2} = t$, we split the integration into an ultraviolet part and an infrared part as follows:
\bea
\frac{1}{6 f^2_\textrm{ind}} &=&  \frac{\pi^2}{13824}(I_{UV} + I_{IR})
\nn\\
I_{\textrm{UV}} &=& \int_{0}^{t_{0}} dt t^{3} \Psi(t)
\nn\\
I_{\textrm{IR}} &=& \int^{\infty}_{t_{0}} dt t^{3} \Psi(t)
\label{UV-IR-rho}
\eea
where
\begin{equation}
\Psi(x) =  \langle T \{{\bar T}(x){\bar T}(0)\}\rangle.
\end{equation}
Now let us use the same approach as in Ref.~\cite{DM} to evaluate $f^2_\textrm{ind}$ which would arise in QCD. We will employ the same ingredients, namely perturbation theory and the operator product  expansion (OPE) at short distances and modern lattice glueball studies at long distances.

Using the same expression for the infrared contribution to $\Psi(x)$ as presented in Ref.~\cite{DM}, one gets the following result for the $I_{\textrm{IR}}$ part:
\beq
I_{\textrm{IR}} = \frac{16 \lambda^2 }{\pi ^2 M_{G}^6}
G_{1,3}^{3,0}\left(\frac{M_{G}^2 t_0}{4} \bigg|
\begin{array}{c}
 1 \\
 0,3,4 \\
\end{array}
\right)
\eeq
where
$$
G_{p,q}^{m,n}\left(z\bigg|
\begin{array}{c}
 a_{1},\ldots,a_{n},\ldots,a_{p} \\
 b_{1},\ldots,b_{m},\ldots,b_{q} \\
\end{array}
\right)
$$
is the Meijer G-function~\cite{Prudnikov:86}, $M_g$ is the glueball mass and
\beq
\lambda = \langle 0| {\bar T}(0)| S\rangle   \ \ ,
\eeq
is the glueball coupling, with $| S\rangle$ being the normalized scalar glueball state.

As for $I_{\textrm{UV}}$ portion, one finds perturbative contributions coming from short-distance scales as well as contributions coming from intermediate energies which are going to be evaluated through the OPE technique. Regarding these latter terms, the results presented in Ref.~\cite{DM} (and references cited therein) allows one to obtain
\beq
I^{OPE}_{UV} = \frac{b^{2}_{0}\alpha_{s}^{2}}{256\pi^{4}}\left\{\frac{2 b_{0} t_0^2}{\pi}  \langle\alpha_{s}F^2\rangle
+ \frac{2 t_0^3}{3} \left[4 + \frac{29 \alpha_{s} }{3}\left(\ln\left(\frac{t_0^3\mu^{6}}{64}\right) + 6 \gamma - 1\right)\right]\langle g F^3\rangle \right\},
\eeq
where $\alpha_{s} = g^{2}/4\pi$ and
$$
b_0 = \frac{11}{3}N_{c} - \frac{2}{3}N_{f},
$$
with $N_{f} = 0$ and $N_{c} = 3$ for gluonic QCD. In addition, $\mu$ is an arbitrary subtraction point, $\gamma =  0.5772$ is the Euler-Mascheroni constant and $\langle (\cdots) \rangle$ are gluon condensate terms
\bea
\langle\alpha_{s}F^2\rangle &=& \langle \alpha_{s}F^{a}_{\mu\nu}F^{a\mu\nu} \rangle
\nn\\
\langle g F^3\rangle &=& \langle g f^{abc}F^{a}_{\mu\nu}F^{b\nu}_{\ \ \rho}F^{c\rho\mu} \rangle
\nn\\
\langle\alpha^2_{s}F^4\rangle &=& 14 \langle (\alpha_{s}f^{abc}F^{a}_{\mu\rho}F^{b}_{\nu}{}^{\rho})^2 \rangle
- \langle (\alpha_{s}f^{abc}F^{a}_{\mu\nu}F^{b}_{\rho\lambda})^2 \rangle
\eea
As for the perturbative contribution, with the change of variables $u = \Lambda_{\textrm{QCD}}^{2} t$ and again using the results quoted in Ref.~\cite{DM}, one gets
\bea
I^{L}_{UV} &=& C_{\Psi} \int_{0}^{u_0} \frac{du}{u}\,\frac{\Theta(u)}{(\ln u)^{2}}
\nn\\
\Theta(u) &=& 1 + \sum_{n=1}^{\infty}\sum_{m = 0}^{n} a_{mn}\,\frac{[\ln(\ln u^{-1})]^{m}}{(\ln u^{-1})^{n}}
\nn\\
C_\Psi &=& \frac{96}{\pi^4}
\eea
where the QCD scale parameter is given by (at one-loop order)
\bea
&& \Lambda_{\textrm{QCD}}(g(\mu),\mu) = \mu e^{-1/[b g^{2}(\mu^2)]}
\nn\\
&& b = \frac{b_0}{8\pi^{2}}.
\eea
The coefficients $a_{mn}$ are loop corrections of higher order. We employ the restriction that $u_{0} = \Lambda_{\textrm{QCD}}^{2} t_{0} < 1$. Let us rewrite $I^{L}_{UV}$ as
\beq
I^{L}_{UV} = C_{\Psi} \int^{\infty}_{\ln u_0^{-1}} dv \frac{\Theta(e^{-v})}{v^{2}}.
\eeq
At leading order the integral can be easily evaluated and the result is
\beq
I^{L}_{UV} = \frac{C_{\Psi}}{x_0}.
\eeq
where $x_0 = \ln u_0^{-1}$. Hence collecting our results, one has that
\bea
\frac{1}{6 f^2_\textrm{ind}} &=&  \frac{\pi^2}{13824}\left\{
\frac{16 \lambda^2 }{\pi ^2 M_{G}^6}
G_{1,3}^{3,0}\left(\frac{M_{G}^2 X_0^2}{4} \bigg|
\begin{array}{c}
 1 \\
 0,3,4 \\
\end{array}
\right)
+ \frac{C_{\Psi}}{\ln[(\Lambda_{\textrm{QCD}} X_{0})^{-2}]}
\right.
\nn\\
&+&\, \left. \frac{b^{2}_{0}\alpha_{s}^{2}}{256\pi^{4}}\left[\frac{2 b_{0} X_0^4}{\pi}
\langle\alpha_{s}F^2\rangle
+ \frac{2 X_0^6}{3} \left[4 + \frac{29 \alpha_{s} }{3}\left(\ln\left(\frac{X_0^6\mu^{6}}{64}\right) + 6 \gamma - 1\right)\right]\langle g F^3\rangle \right]
\right\}.
\eea
The above expression gives the induced $f^2$ as a function of $X_0 = \sqrt{t_0}$. Using the lattice data given by Ref.~\cite{Chen:2005mg}, as well as the values given by Refs.~\cite{Chen:2005mg, Forkel:01, Bagan:1990sy} for the OPE coefficients,  following Ref.~\cite{DM} we quote our result for a matching scale of $X_0^{-1} = 2$~GeV:
\beq
\frac{1}{6 f^2_\textrm{ind}} = 0.00079 \pm 0.00030.
\eeq
The error bar is determined by examining changes in the input parameters. As one can easily see, QCD predicts a positive shift for this coupling constant (in the energy range of interest). This suggests that $f^2_\textrm{ind}$ should be a positive quantity.


\begin{thebibliography}{99}

\bibitem{Einhorn}  M.~B.~Einhorn and D.~R.~T.~Jones,
  ``Naturalness and Dimensional Transmutation in Classically Scale-Invariant Gravity,''
  JHEP {\bf 1503}, 047 (2015)
  [arXiv:1410.8513 [hep-th]].\\
 T.~Jones and M.~Einhorn,
  ``Quantum Gravity and Dimensional Transmutation,''
  PoS PLANCK {\bf 2015}, 061 (2015).

\bibitem{Strumia}
A.~Salvio and A.~Strumia,
  ``Agravity,''
  JHEP {\bf 1406}, 080 (2014)
  [arXiv:1403.4226 [hep-ph]].\\
 A.~Salvio and A.~Strumia,
  ``Agravity up to infinite energy,''
  Eur.\ Phys.\ J.\ C {\bf 78}, no. 2, 124 (2018)
  doi:10.1140/epjc/s10052-018-5588-4
  [arXiv:1705.03896 [hep-th]].

 \bibitem{Donoghue}  J.~F.~Donoghue,
  ``Conformal model of gravitons,''
  Phys.\ Rev.\ D {\bf 96}, no. 4, 044006 (2017)
  [arXiv:1609.03524 [hep-th]].\\
  J.~F.~Donoghue,
  ``Quartic propagators, negative norms and the physical spectrum,''
  Phys.\ Rev.\ D {\bf 96}, no. 4, 044007 (2017)
  [arXiv:1704.01533 [hep-th]].
J.~F.~Donoghue,
  ``Is the spin connection confined or condensed?,''
  Phys.\ Rev.\ D {\bf 96}, no. 4, 044003 (2017)
  [arXiv:1609.03523 [hep-th]].


\bibitem{DM} J.~F.~Donoghue and G.~Menezes,
  ``Inducing the Einstein action in QCD-like theories,''
  Phys.\ Rev.\ D {\bf 97}, 056022 (2018)
  arXiv:1712.04468 [hep-ph].


\bibitem{Holdom}   B.~Holdom and J.~Ren,
  ``QCD analogy for quantum gravity,''
  Phys.\ Rev.\ D {\bf 93}, no. 12, 124030 (2016)
  [arXiv:1512.05305 [hep-th]].\\
  B.~Holdom and J.~Ren,
  ``Quadratic gravity: from weak to strong,''
  arXiv:1605.05006 [hep-th].

\bibitem{Mannheim}
  P.~D.~Mannheim,
  ``Unitarity of loop diagrams for the ghost-like $1/(k^2-M_1^2)-1/(k^2-M_2^2)$ propagator,''
  arXiv:1801.03220 [hep-th].\\
  P.~D.~Mannheim,
  ``Making the Case for Conformal Gravity,''
  Found.\ Phys.\  {\bf 42}, 388 (2012)
  doi:10.1007/s10701-011-9608-6
  [arXiv:1101.2186 [hep-th]].


\bibitem{Hooft}
G.~'t Hooft,
  ``The Conformal Constraint in Canonical Quantum Gravity,''
  arXiv:1011.0061 [gr-qc].\\
 G.~'t Hooft,
  ``Local conformal symmetry: The missing symmetry component for space and time,''
  Int.\ J.\ Mod.\ Phys.\ D {\bf 24}, no. 12, 1543001 (2015).
  doi:10.1142/S0218271815430014
  G.~'t Hooft,
  ``Singularities, horizons, firewalls, and local conformal symmetry,''
  arXiv:1511.04427 [gr-qc].

\bibitem{Shapiro} S.~D.~Odintsov and I.~L.~Shapiro,
  ``General relativity as the low-energy limit in higher derivative quantum gravity,''
  Class.\ Quant.\ Grav.\  {\bf 9}, 873 (1992)
  [Theor.\ Math.\ Phys.\  {\bf 90}, 319 (1992)]
  [Teor.\ Mat.\ Fiz.\  {\bf 90}, 469 (1992)].

\bibitem{Tomboulis}
 E.~T.~Tomboulis,
  ``Renormalization and unitarity in higher derivative and nonlocal gravity theories,''
  Mod.\ Phys.\ Lett.\ A {\bf 30}, no. 03n04, 1540005 (2015).
  doi:10.1142/S0217732315400052\\
E.~Tomboulis,
  ``1/N Expansion and Renormalization in Quantum Gravity,''
  Phys.\ Lett.\  {\bf 70B}, 361 (1977).
  I.~Antoniadis and E.~T.~Tomboulis,
  ``Gauge Invariance and Unitarity in Higher Derivative Quantum Gravity,''
  Phys.\ Rev.\ D {\bf 33}, 2756 (1986).
  doi:10.1103/PhysRevD.33.2756


\bibitem{Smilga}   A.~V.~Smilga,
  ``Ghost-free higher-derivative theory,''
  Phys.\ Lett.\ B {\bf 632}, 433 (2006)
  [hep-th/0503213].\\
 A.~Smilga,
  ``Classical and quantum dynamics of higher-derivative systems,''
  Int.\ J.\ Mod.\ Phys.\ A {\bf 32}, no. 33, 1730025 (2017)
  doi:10.1142/S0217751X17300253
  [arXiv:1710.11538 [hep-th]].

\bibitem{Narain}  G.~Narain and R.~Anishetty,
  ``Short Distance Freedom of Quantum Gravity,''
  Phys.\ Lett.\ B {\bf 711}, 128 (2012).
 G.~Narain,
  ``Signs and Stability in Higher-Derivative Gravity,''
  arXiv:1704.05031 [hep-th].

\bibitem{Ans}   D.~Anselmi,
  ``On the quantum field theory of the gravitational interactions,''
  JHEP {\bf 1706}, 086 (2017)
  doi:10.1007/JHEP06(2017)086
  [arXiv:1704.07728 [hep-th]].\\
 D.~Anselmi,
  ``Fakeons And Lee-Wick Models,''
  arXiv:1801.00915 [hep-th].\\
 D.~Anselmi and M.~Piva,
  ``Perturbative unitarity of Lee-Wick quantum field theory,''
  Phys.\ Rev.\ D {\bf 96}, no. 4, 045009 (2017)
  doi:10.1103/PhysRevD.96.045009
  [arXiv:1703.05563 [hep-th]].

\bibitem{Sobreiro:12}
	R.~F.~Sobreiro, A.~A.~Tomaz and V.~J.~V.~Otoya,
	``de Sitter gauge theories and induced gravities",
	Eur. Phys. J. C {\bf 72}, 1991 (2012)
	doi:10.1140/epjc/s10052-012-1991-4
	[arXiv:1109.0016 [hep-th]].

\bibitem{Stelle:1976gc}
  K.~S.~Stelle,
  ``Renormalization of Higher Derivative Quantum Gravity,''
  Phys.\ Rev.\ D {\bf 16}, 953 (1977).
  doi:10.1103/PhysRevD.16.953

\bibitem{Julve:1978xn}
  J.~Julve and M.~Tonin,
  ``Quantum Gravity with Higher Derivative Terms,''
  Nuovo Cim.\ B {\bf 46}, 137 (1978).
  doi:10.1007/BF02748637

\bibitem{Fradkin:1981hx}
  E.~S.~Fradkin and A.~A.~Tseytlin,
  ``Renormalizable Asymptotically Free Quantum Theory of Gravity,''
  Phys.\ Lett.\  {\bf 104B}, 377 (1981).
  doi:10.1016/0370-2693(81)90702-4

\bibitem{Fradkin:1981iu}
  E.~S.~Fradkin and A.~A.~Tseytlin,
  ``Renormalizable asymptotically free quantum theory of gravity,''
  Nucl.\ Phys.\ B {\bf 201}, 469 (1982).
  doi:10.1016/0550-3213(82)90444-8

\bibitem{Adler:1980pg}
  S.~L.~Adler,
  ``A Formula for the Induced Gravitational Constant,''
  Phys.\ Lett.\  {\bf 95B}, 241 (1980).
  doi:10.1016/0370-2693(80)90478-5

\bibitem{Zee:1980sj}
  A.~Zee,
  ``Spontaneously Generated Gravity,''
  Phys.\ Rev.\ D {\bf 23}, 858 (1981).
  doi:10.1103/PhysRevD.23.858

\bibitem{Zee:1981mk}
  A.~Zee,
  ``Calculating Newton's Gravitational Constant in Infrared Stable {Yang-Mills} Theories,''
  Phys.\ Rev.\ Lett.\  {\bf 48}, 295 (1982).
  doi:10.1103/PhysRevLett.48.295



\bibitem{Adler:1982ri}
  S.~L.~Adler,
  ``Einstein Gravity as a Symmetry Breaking Effect in Quantum Field Theory,''
  Rev.\ Mod.\ Phys.\  {\bf 54}, 729 (1982)
  Erratum: [Rev.\ Mod.\ Phys.\  {\bf 55}, 837 (1983)].
  doi:10.1103/RevModPhys.54.729

\bibitem{Brown:1982am}
  L.~S.~Brown and A.~Zee,
  ``Response To Gravitational Probes And Induced Newton's Constant,''
  Journal Math. Phys. {\bf 24}, 1821 (1983).
  DOE/ER/40048-30 P2.


\bibitem{bos}
I.~L.~Buchbinder, S.~D.~Odintsov and I.~L.~Shapiro,
  ``Effective action in quantum gravity,''
  Bristol, UK: IOP (1992)


\bibitem{Alvarez-Gaume:16}
	L. Alvarez-Gaume, A. Kehagias, C. Kounnas, D.  L\"ust, and A. Riotto,
	``Aspects of quadratic gravity",
	Fortschr. Phys. {\bf 64}, 176 (2016).




\bibitem{Lee:1969fy}
  T.~D.~Lee and G.~C.~Wick,
  ``Negative Metric and the Unitarity of the S Matrix,''
  Nucl.\ Phys.\ B {\bf 9}, 209 (1969).
  doi:10.1016/0550-3213(69)90098-4

\bibitem{Lee:1969fz}
  T.~D.~Lee and G.~C.~Wick,
  ``Unitarity in the $N \theta \theta$ Sector of Soluble Model With Indefinite Metric,''
  Nucl.\ Phys.\ B {\bf 10}, 1 (1969).
  doi:10.1016/0550-3213(69)90275-2


\bibitem{Lee:1970iw}
  T.~D.~Lee and G.~C.~Wick,
  ``Finite Theory of Quantum Electrodynamics,''
  Phys.\ Rev.\ D {\bf 2}, 1033 (1970).
  doi:10.1103/PhysRevD.2.1033


\bibitem{Cutkosky:1969fq}
  R.~E.~Cutkosky, P.~V.~Landshoff, D.~I.~Olive and J.~C.~Polkinghorne,
  ``A non-analytic S matrix,''
  Nucl.\ Phys.\ B {\bf 12}, 281 (1969).
  doi:10.1016/0550-3213(69)90169-2



\bibitem{Grinstein:2007mp}
  B.~Grinstein, D.~O'Connell and M.~B.~Wise,
  ``The Lee-Wick standard model,''
  Phys.\ Rev.\ D {\bf 77}, 025012 (2008)
  doi:10.1103/PhysRevD.77.025012
  [arXiv:0704.1845 [hep-ph]].

\bibitem{Grinstein:2008bg}
  B.~Grinstein, D.~O'Connell and M.~B.~Wise,
  ``Causality as an emergent macroscopic phenomenon: The Lee-Wick O(N) model,''
  Phys.\ Rev.\ D {\bf 79}, 105019 (2009)
  doi:10.1103/PhysRevD.79.105019
  [arXiv:0805.2156 [hep-th]].

\bibitem{Modesto:16}
	L.~Modesto and I.~L.~Shapiro,
	``Superrenormalizable quantum gravity with complex ghosts",
	Phys. Lett. B {\bf 755}, 279 (2016)
	doi:10.1016/j.physletb.2016.02.021
	[arXiv:1512.07600 [hep-th]].

\bibitem{Accioly:17}
	A.~Accioly, B.~L.~Giacchini, and I.~L.~Shapiro,
	``Low-energy effects in a higher-derivative gravity model with real and complex massive poles",
	Phys. Rev. D {\bf 96}, 104004 (2017)
	doi: 10.1103/PhysRevD.96.104004
	[arXiv:1610.05260 [gr-qc]].


\bibitem{Woodard:2015zca}
  R.~P.~Woodard,
  ``Ostrogradsky's theorem on Hamiltonian instability,''
  Scholarpedia {\bf 10}, no. 8, 32243 (2015)
  doi:10.4249/scholarpedia.32243
  [arXiv:1506.02210 [hep-th]].

\bibitem{Bender:2007wu}
  C.~M.~Bender and P.~D.~Mannheim,
  ``No-ghost theorem for the fourth-order derivative Pais-Uhlenbeck oscillator model,''
  Phys.\ Rev.\ Lett.\  {\bf 100}, 110402 (2008)
  doi:10.1103/PhysRevLett.100.110402
  [arXiv:0706.0207 [hep-th]].



\bibitem{Salvio:2015gsi}
  A.~Salvio and A.~Strumia,
  ``Quantum mechanics of 4-derivative theories,''
  Eur.\ Phys.\ J.\ C {\bf 76}, no. 4, 227 (2016)
  doi:10.1140/epjc/s10052-016-4079-8
  [arXiv:1512.01237 [hep-th]].

\bibitem{Raidal:2016wop}
  M.~Raidal and H.~Veermäe,
  Nucl.\ Phys.\ B {\bf 916}, 607 (2017)
  doi:10.1016/j.nuclphysb.2017.01.024
  [arXiv:1611.03498 [hep-th]].


\bibitem{Kallen:1952zz}
  G.~Kallen,
  ``On the definition of the Renormalization Constants in Quantum Electrodynamics,''
  Helv.\ Phys.\ Acta {\bf 25}, no. 4, 417 (1952).
  doi:10.1007/978-3-319-00627-7-90


\bibitem{Lehmann:1954xi}
  H.~Lehmann,
  ``On the Properties of propagation functions and renormalization contants of quantized fields,''
  Nuovo Cim.\  {\bf 11}, 342 (1954).
  doi:10.1007/BF02783624


\bibitem{Oehme:1979ai}
  R.~Oehme and W.~Zimmermann,
  ``Quark and Gluon Propagators in Quantum Chromodynamics,''
  Phys.\ Rev.\ D {\bf 21}, 471 (1980).
  doi:10.1103/PhysRevD.21.471

\bibitem{Cornwall:2013zra}
  J.~M.~Cornwall,
  ``Positivity violations in QCD,''
  Mod.\ Phys.\ Lett.\ A {\bf 28}, 1330035 (2013)
  doi:10.1142/S0217732313300358
  [arXiv:1310.7897 [hep-ph]].




\bibitem{deBerredoPeixoto:2003pj}
  G.~de Berredo-Peixoto and I.~L.~Shapiro,
  ``Conformal quantum gravity with the Gauss-Bonnet term,''
  Phys.\ Rev.\ D {\bf 70}, 044024 (2004)
  doi:10.1103/PhysRevD.70.044024
  [hep-th/0307030].






\bibitem{Avramidi:1985ki}
  I.~G.~Avramidi and A.~O.~Barvinsky,
  ``Asymptotic Freedom In Higher Derivative Quantum Gravity,''
  Phys.\ Lett.\  {\bf 159B}, 269 (1985).
  doi:10.1016/0370-2693(85)90248-5

\bibitem{Avramidi:1986mj}
  I.~G.~Avramidi,
  ``Covariant methods for the calculation of the effective action in quantum field theory and investigation of higher derivative quantum gravity,''
  hep-th/9510140.



\bibitem{Zee:1981ff}
  A.~Zee,
  ``A Theory of Gravity Based on the {Weyl-Eddington} Action,''
  Phys.\ Lett.\  {\bf 109B}, 183 (1982).
  doi:10.1016/0370-2693(82)90749-3

\bibitem{Zee:1983mj}
  A.~Zee,
  ``Einstein Gravity Emerging From Quantum Weyl Gravity,''
  Annals Phys.\  {\bf 151} (1983) 431.
  doi:10.1016/0003-4916(83)90286-5


\bibitem{'tHV}
G.~'t Hooft and M.~J.~G.~Veltman,
``One Loop Divergencies in the Theory of Gravitation,''
Annales Poincare Phys.\ Theor.\ A {\bf 20} (1974) 69.

\bibitem{Donoghue:1994dn}
  J.~F.~Donoghue,
  ``General Relativity As An Effective Field Theory: The Leading Quantum
  Corrections,''
  Phys.\ Rev.\  D {\bf 50}, 3874 (1994)
  [arXiv:gr-qc/9405057].



\bibitem{Dona:2015tra}
  P.~Donà, S.~Giaccari, L.~Modesto, L.~Rachwal and Y.~Zhu,
  ``Scattering amplitudes in super-renormalizable gravity,''
  JHEP {\bf 1508}, 038 (2015)
  doi:10.1007/JHEP08(2015)038
  [arXiv:1506.04589 [hep-th]].

\bibitem{Johansson:2017srf}
  H.~Johansson and J.~Nohle,
  ``Conformal Gravity from Gauge Theory,''
  arXiv:1707.02965 [hep-th].

\bibitem{Maldacena:2011mk}
  J.~Maldacena,
  ``Einstein Gravity from Conformal Gravity,''
  arXiv:1105.5632 [hep-th].

\bibitem{Adamo:2013tja}
  T.~Adamo and L.~Mason,
  ``Conformal and Einstein gravity from twistor actions,''
  Class.\ Quant.\ Grav.\  {\bf 31}, no. 4, 045014 (2014)
  doi:10.1088/0264-9381/31/4/045014
  [arXiv:1307.5043 [hep-th]].

\bibitem{Adamo:2016ple}
  T.~Adamo, P.~Hhnel and T.~McLoughlin,
  ``Conformal higher spin scattering amplitudes from twistor space,''
  JHEP {\bf 1704}, 021 (2017)
  doi:10.1007/JHEP04(2017)021
  [arXiv:1611.06200 [hep-th]].


\bibitem{Han}
	T. Han and S. Willenbrock,
	``Scale of quantum gravity",
	Phys. Lett. B {\bf 616}, 215 (2005).


\bibitem{Aydemir}
	U. Aydemir, M. M. Anber, and J. F. Donoghue,
	``Self-healing of unitarity in effective field theories and the onset of new physics",
	Phys. Rev. D {\bf 86}, 014025 (2012).


\bibitem{Holdom:2007gg}
  B.~Holdom,
  ``Does Massless QCD has vacuum energy?,''
  New J.\ Phys.\  {\bf 10}, 053040 (2008)
  doi:10.1088/1367-2630/10/5/053040
  [arXiv:0708.1057 [hep-ph]].\\
   B.~Holdom,
  ``Mass gap without vacuum energy,''
  Phys.\ Lett.\ B {\bf 681}, 287 (2009)
  doi:10.1016/j.physletb.2009.10.021
  [arXiv:0907.0009 [hep-ph]].


\bibitem{Caswell:1974gg}
  W.~E.~Caswell,
  ``Asymptotic Behavior of Nonabelian Gauge Theories to Two Loop Order,''
  Phys.\ Rev.\ Lett.\  {\bf 33}, 244 (1974).
  doi:10.1103/PhysRevLett.33.244


\bibitem{Banks:1981nn}
  T.~Banks and A.~Zaks,
  ``On the Phase Structure of Vector-Like Gauge Theories with Massless Fermions,''
  Nucl.\ Phys.\ B {\bf 196}, 189 (1982).
  doi:10.1016/0550-3213(82)90035-9



\bibitem{Anderson:1971pn}
  J.~L.~Anderson and D.~Finkelstein,
  ``Cosmological constant and fundamental length,''
  Am.\ J.\ Phys.\  {\bf 39}, 901 (1971).
  doi:10.1119/1.1986321

\bibitem{vanderBij:1981ym}
  J.~J.~van der Bij, H.~van Dam and Y.~J.~Ng,
  ``The Exchange of Massless Spin Two Particles,''
  Physica {\bf 116A}, 307 (1982).
  doi:10.1016/0378-4371(82)90247-3


\bibitem{Buchmuller:1988wx}
  W.~Buchmuller and N.~Dragon,
  ``Einstein Gravity From Restricted Coordinate Invariance,''
  Phys.\ Lett.\ B {\bf 207}, 292 (1988).
  doi:10.1016/0370-2693(88)90577-1

\bibitem{Unruh:1988in}
  W.~G.~Unruh,
  ``A Unimodular Theory of Canonical Quantum Gravity,''
  Phys.\ Rev.\ D {\bf 40}, 1048 (1989).
  doi:10.1103/PhysRevD.40.1048

\bibitem{Weinberg:1988cp}
  S.~Weinberg,
  ``The Cosmological Constant Problem,''
  Rev.\ Mod.\ Phys.\  {\bf 61}, 1 (1989).
  doi:10.1103/RevModPhys.61.1

\bibitem{Henneaux:1989zc}
  M.~Henneaux and C.~Teitelboim,
  ``The Cosmological Constant and General Covariance,''
  Phys.\ Lett.\ B {\bf 222}, 195 (1989).
  doi:10.1016/0370-2693(89)91251-3


\bibitem{Percacci:2017fsy}
R. Percacci,
  ``Unimodular quantum gravity and the cosmological constant,''
  arXiv:1712.09903 [gr-qc].


\bibitem{Prudnikov:86}
A.~P.~Prudnikov, Yu~A.~Brychkov and O.~I.~Marichev,
{\it Integrals and Series}, vol. 2
(Gordon and Breach, London, 1986).

\bibitem{Chen:2005mg}
  Y.~Chen {\it et al.},
  ``Glueball spectrum and matrix elements on anisotropic lattices,''
  Phys.\ Rev.\ D {\bf 73}, 014516 (2006)
  [hep-lat/0510074].

\bibitem{Forkel:01}
H. Forkel,
``Scalar gluonium and instantons",
 Phys. Rev. D {\bf 64}, 034015 (2001).


\bibitem{Bagan:1990sy}
  E.~Bagan and T.~G.~Steele,
  ``Mass of the Scalar Glueball: Higher Loop Effects in the {QCD} Sum Rules,''
  Phys.\ Lett.\ B {\bf 243}, 413 (1990).




\end{thebibliography}
\end{document}